\documentclass[10pt,twocolumn,twoside]{IEEEtran}

\usepackage{cite}
\usepackage{amsmath}
\usepackage{amssymb}
\usepackage{fontenc}
\usepackage{color}

\newtheorem{theorem}{Theorem}

\newtheorem{remark}{Remark}
\usepackage{algorithm}
\usepackage{algorithmic}
\usepackage{subfigure}
\usepackage{graphicx}
\usepackage{graphicx}
\usepackage[caption=false,font=footnotesize]{subfig}
\usepackage{url}
\usepackage{booktabs}
\usepackage{balance}
\usepackage{epstopdf}
\usepackage{xcolor}

\newcommand{\tabincell}[2]{\begin{tabular}{@{}#1@{}}#2\end{tabular}}

\begin{document}

\title{Hybrid Policy Learning for Energy-Latency Tradeoff in MEC-Assisted VR Video Service}

\author{Chong~Zheng,~\IEEEmembership{Student~Member,~IEEE},
        Shengheng~Liu,~\IEEEmembership{Member,~IEEE},
        Yongming~Huang,~\IEEEmembership{Senior Member,~IEEE},
        Luxi~Yang,~\IEEEmembership{Senior Member,~IEEE}
\vspace{-2em}

\thanks{Manuscript received October 16, 2020; revised XXX XX, XXXX; accepted XXX XX, XXXX. Date of publication XXX XX, XXXX; date of current version XXX XX, XXXX. This work was supported in part by the National Natural Science Foundation of China under Grant Nos. 62001103 and 61720106003, and the National Key R\&D Program of China under Grant No. 2018YFB1800801. Part of this work was presented at the IEEE 18$^{\rm{th}}$ Wireless Communications and Networking Conference (WCNC), Seoul, South Korea, May 2020 \cite{Zheng20}. (Corresponding author: Y.~Huang.)}
\thanks{The authors are with the School of Information Science and Engineering, Southeast University, Nanjing 210096, China, and also with the Purple Mountain Laboratories, Nanjing 211111, China (e-mail: \{czheng; s.liu; huangym; lxyang\}@seu.edu.cn).}
}

\markboth{IEEE TRANSACTIONS ON VEHICULAR TECHNOLOGY,~Vol.~XX, No.~X, XXX~2021}%
{Zheng \MakeLowercase{\textit{et al.}}: HYBRID POLICY LEARNING FOR ENERGY-LATENCY TRADEOFF IN MEC-ASSISTED VR VIDEO SERVICE}

\maketitle

\begin{abstract}
Virtual reality (VR) is promising to fundamentally transform a broad spectrum of industry sectors and the way humans interact with virtual content. However, despite unprecedented progress, current networking and computing infrastructures are incompetent to unlock VR's full potential. In this paper, we consider delivering the wireless multi-tile VR video service over a mobile edge computing (MEC) network. The primary goal is to minimize the system latency/energy consumption and to arrive at a tradeoff thereof. To this end, we first cast the time-varying view popularity as a model-free Markov chain to effectively capture its dynamic characteristics. After jointly assessing the caching and computing capacities on both the MEC server and the VR playback device, a hybrid policy is then implemented to coordinate the dynamic caching replacement and the deterministic offloading, so as to fully utilize the system resources. The underlying multi-objective problem is reformulated as a partially observable Markov decision process, and a deep deterministic policy gradient algorithm is proposed to iteratively learn its solution, where a long short-term memory neural network is embedded to continuously predict the dynamics of the unobservable popularity. Simulation results demonstrate the superiority of the proposed scheme in achieving a trade-off between the energy efficiency and the latency reduction over the baseline methods.
\end{abstract}


\vspace{-1em}

\section{Introduction}

\IEEEPARstart{T}{he} past few years have witnessed the rapid evolution of virtual reality (VR), which has branched out into numerous application domains, from education to healthcare to the alluring world of entertainment. VR implies immersing users inside a synthetic fictional world where the physical environment and interactions are simulated. The ultimate goal of VR interface design is to break the barrier that separates both worlds by being unable to distinguish between the real world and a simulated one. A crucial step in this direction is to boost the resolution of the VR system to the full fidelity of human perception and to free the user from any cable connection that limits mobility. This requires an intensive computation and transmission of ultra-massive volume of data. For mobile VR devices that suffers from limited battery life and insufficient computing capabilities, the resultant resource request can be highly prohibitive. We thus envision that the next steps toward the future interconnected VR service will come from a flexible use of caching, computing, and connectivity resources. To realize this vision, innovations in the supporting infrastructure are essential, and many trade-offs need to be studied.

\vspace{-1em}
\subsection{Related Works}

From academia to industry, mobile edge computing (MEC) is commonly viewed as a key enabler of seamless and immersive VR experience. It addresses the above technical bottlenecks by implementing the storage, processing, and the service delivery at the wireless network edge, so as to reduce bandwidth demand for backhaul to the backbone network. Many related works on the MEC-based solution can be found in the literatures. For instance, in \cite{Chak17}, the delivery of VR video is considered in a cooperative multi-node MEC system where all the nodes share their resources by backhual links. The optimal strategies are developed to maximize the aggregate reward of the small base stations (SBSs). In \cite{Mang17}, a delivery strategy between the servers and the subscribers is further developed to optimize the bandwidth and latency of the wireless VR video service. Aside from the studies on content delivery, a task scheduling strategy is proposed for the VR video service within a queueing MEC framework \cite{8319985}, which selects the computation model for the MEC server to minimize the communication-resource consumption under the delay constraint. All the devices in the system are assumed to be cache-fixed. Later on, the restriction on the fixed cache is further relaxed. Based on the analysis of the VR video production flow in \cite{Mang17}, an additional pre-caching procedure is employed in \cite{8422519, 8728029}, where partial field of views (FoVs) are cached at the local VR device in advance and certain post-processing tasks can be performed on demand either at the local or the MEC server. Under a similar framework, the mmWave 802.11ad wireless technology is  incorporated into the MEC network \cite{8513999} to promise the use of high-bandwidth wireless VR video service. Also, an adaptive computing offloading scheme, which focuses on the scenario of indoor environment, is designed in \cite{8513999}. For the VR video service in the outdoor environment, a multi-connectivity-based millimeter wave MEC network is developed in \cite{8492440}, and the trade-off among link adaptation, viewpoint rendering offloading, and chunk quality adaptation is sought to improve the utilization of bandwidth and energy.

We observe that the existence of the viewpoints' popularity and the rich caching resources in the MEC system have motivated the commonly-adopted caching strategy. However, all the caching placements in the existing studies are performed in advance and remains constant during the service. As such, the new tiles received for current viewing will be outright discarded, which leads to enormous waste and repetitive transmission in the wireless network because of the FoV overlapping phenomenon \cite{Chen2019Data}. Transmission resource allocation in the up/down link is thus considered in \cite{Chen2019Data}, while the caching and computing power at the edge is, however, omitted. In this context, a dynamic caching replacement scheme is designed in this work to significantly improve the utilization of cache resources and to reduce the waste caused by the redundant transmission. Unlike the existing works where the caching status remains unchanged during the service, the proposed caching policy updates the caching status as soon as new tiles enter and attempts to solve the following technical challenges. On the one hand, in order to effectively realize dynamic and real-time caching replacement, the future view requests must be acquired in advance to support the caching update decision. However, the view request is time-varying and depends on the users' interests, which is unavailable in advance and difficult to accurately predict. On the other hand, real-time caching replacement involves both storage of the incoming prevalent content and removal of the obsolete ones. As such, the decision space is huge with its dimension varies constantly due to the uncertainty of the incoming content.

The abundant computing resources in close proximity to the VR service subscribers provided by the MEC server can be another accelerator of the VR video service. Many computation offloading schemes of the VR video service have been designed to exploit the computing resource and improve the quality of service. For example, unmanned aerial vehicles are introduced to the wireless VR video service in \cite{ChenM19} to provide support in terms of computation, caching, and transmission. To maximize the users' reliability, a distributed deep learning algorithm is adopted to solve the joint caching and transmission optimization problem. In \cite{He18}, a more complicated computation offloading scheme is considered for the VR applications in a multi-user edge computing system. The mobile users' energy minimization is formulated as a convex problem subject to the energy, computing latency, and computing frequency constraints, which is solved by using a classical Lagrangian duality. To maximize the average tolerant delay while guaranteeing a given transmission rate constraint, a joint communication, caching and computing decision problem is considered in \cite{Dang19}, which is converted to a multiple choice multiple dimensional knapsack problem. However, the offloaded tasks in \cite{ChenM19, Dang19, 8319985} are treated as unities and offloaded in a integral form either to local VR device or the MEC server. As a result, idle computing resources are constantly released.

Fortunately, the videos are usually sliced into multiple tiles in wireless VR video service, which facilitates implementation of task segmentation. In the light of this fact, a multi-tile deterministic offloading scheme is proposed in this work, which not only effectively avoids the release of idle computing resources but also fully exploited the cache resources in each device. The proposed offloading mechanism is completely different from the integral offloading in \cite{ChenM19, Dang19, 8319985} or the generic proportional offloading scheme adopted in some MEC studies. Despite remarkable benefits in terms of computing/caching resource utilization, such micromanagement for each single tile results in a dimension explosion of the decision space, which is an important challenge to address in this work. Collectively, caching and offloading policies constitute a hybrid policy that is considered for the intelligent wireless VR video service in this work. The tremendous and time-varying dimension of the hybrid policy space represents a paramount factor to consider, which obviously overwhelms the capacity of the conventional optimization methods. Furthermore, as the computing task in each device is supported by the co-located cached tiles, the present caching status has an important impact on the offloading decision. Conversely, the present offloading decision determines the tile input for each device and consequently affects the caching policy as well as the future caching status. This complicated interaction between the two functions further compounds the problem and highlights the necessity of exploration of more intelligent algorithms.

Apart from exploring the potential of MEC in terms of caching and computing resources, the success of wireless VR video service also requires making full use of the multi-perspective characteristics of VR videos. The multiple views in VR video service allow users to watch the FoV interested in. This is fundamentally different from the ordinary two-dimensional (2D) single-view video service where users can only enjoy the pre-determined contents passively. Taking into account the FoV overlap due to the multi-view, the optimal multi-cast schemes of tiled VR video from one server to multiple wireless users are proposed in \cite{Guo2018Optimal, Guo2019Optimal}, where the minimization of transmission power and the maximization of received video's quality are considered. However, the dynamics of the popularity is neglected and the distribution parameters of the popularity is assumed known in both of these studies. The dynamics in the view popularity is considered in \cite{8395443, Chen2019Data}, where learning-based resources management for uplink/downlink transmissions is studied within small cell networks. However, all the possible distribution parameters of the popularity are still given as \emph{a priori} knowledge during the service. An accurate information of users' popularity can significantly instruct the learning and optimization of the caching policy. However, in reality, the dynamic view popularity as a time-varying characteristics depends on the changes of users' interests and is unavailable in advance. For more realistic modeling, we capture the variations of the popularity using a model-free Markov decision process (MDP). In such a model, no \emph{a priori} knowledge on the popularity is assumed aside from the distribution type, which is more robust yet more difficult to accurately predict. Hereby lies another important technical contribution of this work.

We know from the foregoing discussion that, a large number of parameters and factors can, and should, be considered in optimizing resource allocation and network performance of the collaborative network for wireless VR service. It is also imperative for the hybrid policy optimization to remain flexible and responsive to changes and unpredictable elements in the network conditions. This requires assistance of intelligent methodology. Artificial Intelligence has unleashed a new era of ingenuity. There has been a consensus among global top telecommunications companies such as Huawei, Samsung, and Qualcomm \cite{huawei18, Sam18, qual16} that artificial-intelligence-based end-edge-cloud orchestrated network is the recipe for the commercially viable wireless VR. However, as higher requirements are posed on the computing power and latency of communication services, the intelligent algorithms cannot be simply used as is, without any modification or customization. In the sequel, our technical contributions in terms of system modelling, problem formulation, and algorithm redesign to cope with realistic scenarios will be showcased.

\subsection{Main Contributions}
In this paper, a joint dynamic caching replacement and deterministic offloading scheme for MEC-assisted wireless VR video service is proposed to minimize the service latency as well as the system energy consumption, which achieves a trade-off among the supporting resources in the system. Concretely, the technical contributions of this work include:

\begin{itemize}
  \item Multi-tile deterministic offloading is proposed in this work, which is completely different from the proportional offloading commonly adopted in the generic MEC researches. In the designed offloading policy, we segment the FoV tasks into tiles and assign the computation task of each tile to a certain device as per the hybrid policy and, thereby realize the full linkage and utilization of multi-dimensional resources.

  \item In view of the FoV overlapping, a dynamic caching replacement scheme is designed and operates collaboratively with the deterministic offloading. As such, the repetitive communication overhead is significantly reduced, and the resource utilization at both the MEC server and the local VR device is dramatically improved.

  \item We take into account the dynamics of the viewpoint popularity and first model the variations of the distribution parameter as a completely model-free Markov process, where both the transition probabilities and the distribution parameters are assumed unknown to the system during the service. This model is more closely aligned with reality and has not been studied in the related works.

  \item We convert the underlying optimization problem into a partially observable model-free MDP whose state transition probability is unaccessible and the state space is partially observable. Then, we incorporates the long short-term memory (LSTM) network into the deep deterministic policy gradient (DDPG) algorithm to overcome the deficiency of the original DDPG in capturing the temporal information hidden in the input data. The proposed scheme can also handle large state/action space.
\end{itemize}

The rest of this paper is organized as follow. The system model is presented in Section \ref{sec2}. And the problem is formulated in Section \ref{sec3}. In Section \ref{sec4}, we propose a deep reinforcement learning algorithm to get the hybrid policy for wireless VR video service. In Section \ref{sec5}, simulation results are discussed. Finally, conclusions are drawn in Section \ref{sec6}.

\section{SYSTEM MODEL}
\label{sec2}

\begin{figure}[!pht]
\centering
\includegraphics[width=0.36\textwidth]{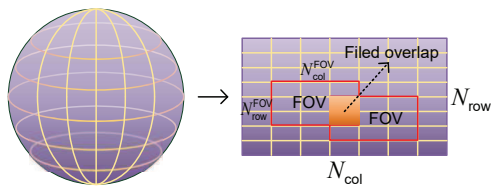}
\caption{Illustration of tiling technique for VR video processing.}
\label{fig1}
\end{figure}

The tiling technique \cite{8728029, Jiang20} illustrated in Fig. \ref{fig1} is adopted to process the VR videos. This technique unfolds the spherical video into a 2D version and further divides the 2D video into pieces of tiles. At each time instant, only the tiles in the FoV that the user focus on will be extracted to dispose rather than the entire spherical video. Thereby, the processed data size is significantly reduced. Concretely, the spherical video on the left-hand side of Fig. \ref{fig1} is converted to the 2D tiles on the right-hand side. Let us take a closer look at the 2D tiles in Fig. \ref{fig1}. The 2D VR video on the righthand side is divided into $N = {N_{{\rm{row}}}} \times {N_{{\rm{col}}}}$ rectangular segments of the same size, where ${N_{{\rm{row}}}}$ and ${N_{{\rm{col}}}}$ respectively denote the numbers of segments in each row and each column. These rectangular segments are referred to as tiles. One FoV is represented by a rectangle with $N_{{\rm{col}}}^{{\rm{FOV}}}$ tiles in column and $N_{{\rm{row}}}^{{\rm{FOV}}}$ tiles in row, and two FoVs can overlap with each other as marked by the orange area. Suppose that one 3D FoV is requested by the user, the related rectangular regions in the 2D plane are extracted immediately, which is computationally simple. The computation task considered in this paper is mainly the conversion from the 2D tiles to the 3D FoV, i.e., 2D tiles are the input of the computation while the 3D FoV file is the output. In addition, we assume that all the tiles can be accessed from the cloud but the caching capabilities of the MEC server and the local VR device are both limited. As such, the VR video can only be partially cached. Suppose that certain tiles are requested whereas neither local VR device nor MEC server has a copy, the MEC server will in turn try to access the tiles from the cloud. For each task, dynamic caching replacement at both the local VR device and the MEC server is considered, and the deterministic offloading of multiple tiles is taken into consideration in this work.

\begin{figure}[!pht]
\centering
\includegraphics[width=0.48\textwidth]{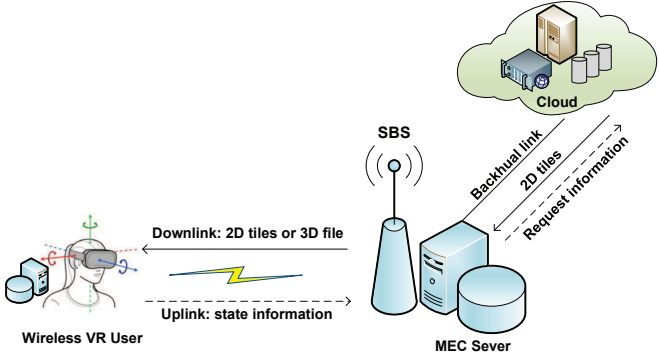}
\caption{Architecture of an MEC-based wireless VR video system.}
\label{fig2}
\end{figure}

As illustrated in Fig. \ref{fig2}, we consider a typical and simple wireless VR video service system with a single VR user wirelessly connected to a small base station (SBS) which is equipped with one MEC server. We assume that both sides are endowed with certain caching and computing capabilities. The MEC server connects with the cloud by a backhual link. When the user watches a VR video, the local VR device generates viewpoint request according to user's head movement. During the service, we assume that user makes a new request as soon as the previous request is satisfied \cite{Chen2019Data}. The interval between two adjacent requests is defined as a time slot. We readily see that the duration of each time slot is not equal due to the different service delays of different tasks. At the beginning of each time slot, the state and the request information are uploaded to the MEC server. Next, the computation task is extracted and divided into several sub-tasks as per the deterministic offloading model proposed in Section \ref{ss:OM}, where all the sub-tasks are assigned to MEC server or local VR device in a smart manner for subsequent computing. Then, both the demanded but absent 2D tiles by the local VR device as well as the outputted 3D FoV file of the MEC server are sent to the user. In certain time slots, the MEC server might also send request to the cloud through the backhual link for some necessary but absent tiles in accomplishing the computation task. In each time slots, the dynamic caching policy is implemented in both VR device and MEC server,  which will be elaborated in section \ref{ss:DCRM}. After all the sub-tasks are completed and all the outputs are acquired by the local VR device, the user's request in the present time slot is considered satisfied.

\subsection{View Request Model}

Let ${\cal V} = \{ {V_k}\}_{k=1}^K$ and ${\cal F} = \{ {F_n}\}_{n=1}^N$ respectively denote the spaces of all the views and tiles in the 2D domain. As shown in Fig. \ref{fig1}, the relationship between the number of viewpoints $K$ and the segmentation of tiles can be expressed as $K = (({N_{{\rm{col}}}} - N_{{\rm{col}}}^{{\rm{FOV}}})/{\Delta _{\rm{h}}} + 1) \times (({N_{{\rm{row}}}} - N_{{\rm{row}}}^{{\rm{FOV}}})/{\Delta _{\rm{v}}} + 1)$. Similar to \cite{Guo2018Optimal}, we let ${\Delta _{\rm{h}}}$ and ${\Delta _{\rm{v}}}$ represent the number of interval tiles for two adjacent views in the horizontal and the vertical directions, respectively. The data size (in bit) of each tile is given by $\tau  = {Q \mathord{\left/ {\vphantom {Q N}} \right.\kern-\nulldelimiterspace} N}$, where $Q$ is the full size of all the 2D tiles. For each ${V_k}$, we assume that it consists of constant tiles, which is in line with the characteristics of human vision and also widely adopted in literatures such as \cite{Chen2019Data, Guo2018Optimal, Guo2019Optimal}.

We denote with ${{\cal F}_{{V_k}}}{\rm{ = \{ }}{F_{{k_1}}}{\rm{,}} \cdots ,{F_{{k_Z}}}{\rm{\} }}$ the space of tiles consistuting the view ${V_k}$, where $Z$ is a constant. For each ${V_k}$, its 3D FoV is computed from the 2D FoV and the input is the set ${{\cal F}_{{V_k}}}$. We let $({D^{{\rm{in}}}},{D^{{\rm{out}}}})$ represent the input and the output of the 2D-to-3D FoV conversion. Typically, in order to create a stereoscopic vision, $\varphi {\rm{ = }}\frac{{{D^{{\rm{out}}}}}}{{{D^{{\rm{in}}}}}} \ge 2$ should be satisfied \cite{8728029,Reichelt2010Depth,Xun2011Converting}. Suppose that in a time slot $t$, the request probability for viewpoint ${V_k} \in {\cal V}$ at the local VR device is ${p_k}(t)$. Additionally, we let ${\bf{p}}(t) = [{p_1}(t), \cdots ,{p_K}(t)]$ denote the popularity of all viewpoint. Thus, we have $\sum_{k = 1}^K {{p_k}(t)}  = 1$. A detailed discussion on the dynamic view popularity is presented in the following subsection.

\subsection{Dynamic Popularity Model}

Viewpoint popularity of the VR video is the external manifestation of the user's subjective interest. Though the changes of popularity is complicated, its statistical representation can be established given a large number of users' traces. Inspired by \cite{Sadeghi2017Optimal} which describes the content popularity of a ordinary caching task as a Markov chain model, we model the dynamic popularity in the VR video service as a completely model-free MDP whose distribution parameters and transition probabilities are both unknown. 
The popularity of the user's request in each time slot is assumed to follow the Zipf distribution \cite{LiJ16,Yang20}, and the distribution parameter ${\gamma _t}$ evolves dynamically over time. As such, the probability of the $k$-th viewpoint in time slot $t$ is
\begin{equation} \label{e1}
p_{k}^{\gamma_{t}}\left(t\right)=\frac{1}{k^{\gamma_{t}}\sum_{l=1}^{K}\frac{1}{l^{\gamma_{t}}}}.
\end{equation}

\begin{figure}[!pht]
\centering
\includegraphics[width=0.31\textwidth]{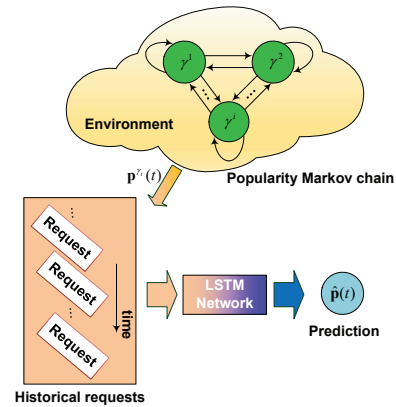}
\caption{Dynamic popularity model and the schematic of its prediction.}
\label{fig3}
\end{figure}

As depicted in Fig. \ref{fig3}, we model the time-varying dynamics of ${\gamma _t}$ using a model-free Markov process with $\left| {\cal G} \right|$ states recorded in the set ${\cal G} = \{ {\gamma ^i}|i = 1,2, \cdots \}$. It is worth mentioning that, in this model, both the transition probabilities and the parameter space are unknown to the system during the VR video service, which is more realistic. The system only has the \emph{a priori} knowledge on the user's historical requests, which is used to assist the prediction of the time-varying view probability. Specifically, as shown in Fig. \ref{fig3}, if the SBS receives a viewpoint request in time slot $t$ under the popularity ${{\bf{p}}^{{\gamma _t}}}(t)$, i.e., ${V_k}(t){|_{{{\bf{p}}^{{\gamma _t}}}(t)}}$, which is abbreviated as ${V_k}(t)$ for the convenience of discussion, we have
\begin{equation} \label{e2}
{\bf{\hat p}}(t) = f({\bf{R}}(t - 1)),
\end{equation}
where ${\bf{R}}(t-1) = [{V_k}(t - {T_r}), \cdots ,{V_k}(t - 1)]$ is the historical request of continuous ${T_r}$ time slots before time slot $t$. ${\bf{\hat p}}(t)$ is the prediction of ${{\bf{p}}^{{\gamma _t}}}(t)$. The function $f( \cdot )$ is a predictive function realized by LSTM, whose implementation will be detailed in section \ref{sec4}. ${\bf{\hat p}}(t)$ is the prediction of ${{\bf{p}}^{{\gamma _t}}}(t)$.

\subsection{Deterministic Task Segmentation and Offloading Model}
\label{ss:OM}

As mentioned above, unlike existing related works in MEC-assisted VR system, deterministic task segmentation and offloading is implemented in this work. This means that rather than processed as a unity, certain particular tiles in the computation input set are selected and computed. Furthermore, as will be explained later, the adopted MEC task offloading scheme also differs from a generic paradigm where only roughly how much proportion of a task should be offloaded to the MEC server is considered.

Suppose that the SBS receives a viewpoint request ${V_k}(t)$ and ${{\cal F}_{{V_k}(t)}}{\rm{ = \{ }}{F_{{k_1}(t)}}{\rm{,}} \cdots ,{F_{{k_Z}(t)}}{\rm{\} }}$. Let ${\bf{o}}(t){\rm{ = [}}{{\rm{o}}_1}{\rm{(t),}} \cdots ,{{\rm{o}}_Z}{\rm{(t)]}}$ denote the $1 \times Z$ binary offloading action vector in slot $t$, where ${{\rm{o}}_z}(t) = 1$ means that the tile ${F_{{k_z}(t)}}$ is computed at the MEC server and otherwise it is computed in the local VR device. Particularly, $\sum\nolimits_{z = 1}^Z {{{\rm{o}}_z}(t) = Z}$ indicates that all the tiles will be computed at MEC server, and the opposite is true if $\sum\nolimits_{z = 1}^Z {{{\rm{o}}_z}(t) = 0}$. For the convenience of discussion, we let ${{\cal F}_{\rm{M}}}(t)={\rm{\{ }}{F_{k_1^{\rm{M}}(t)}}{\rm{,}} \cdots ,{F_{k_{S(t)}^{\rm{M}}(t)}}{\rm{\} }} \subseteq {{\cal F}_{{V_k}(t)}}$  and ${{\cal F}_{\rm{L}}}(t) = {\rm{\{ }}{F_{k_1^{\rm{L}}(t)}}{\rm{,}} \cdots ,{F_{k_{R(t)}^{\rm{L}}(t)}}{\rm{\} }} \subseteq {{\cal F}_{{V_k}(t)}}$ respectively denote the tiles sets computed in MEC server and local VR device as per the offloading action ${\bf{o}}(t)$. Note that $S(t)$ and $R(t)$ can be different in each slot. But $S(t) + R(t) = Z$ and ${{\cal F}_{\rm{M}}(t) \cap {{\cal F}_{\rm{L}}}}(t) = \emptyset$ should be always satisfied.

\begin{remark}\label{remark0}
The sliced input files produced by tiling technique facilitates the segmentation of task. In addition, the constraint ${{\cal F}_{\rm{M}}(t) \cap {{\cal F}_{\rm{L}}}}(t) = \emptyset$ ensures that the two computing outputs are non-overlapping, such that the outputs on both devices are splice easy.
\end{remark}

\subsection{Computation Model}

The MEC server is assumed to run at a given computational frequency denoted as ${f_{\rm{M}}}$, which is related to the central processing unit (CPU) specification. The computation task in wireless VR service mainly comprises the projection and rendering of the 2D tiles to the requested 3D FoV. Such operation is commonly deemed highly computation-intensive and, thus, the computation time is considered to have a linear relation with respect to the computational frequency \cite{Mang17, 8728029} Therefore, for request ${V_k}(t)$ under ${\bf{o}}(t)$ the computation latency at the MEC server is
\begin{equation} \label{e3}
T_{{\rm com}}^{{\rm M}}(t)=\left.\left(w\tau\sum\nolimits _{z=1}^{Z}o_{z}(t)\right)\right/f_{{\rm M}},
\end{equation}
where $w$ is the CPU cycles involved in computing one-bit data. The computation latency at local VR device can be given by
\begin{equation} \label{e4}
T_{{\rm com}}^{{\rm L}}(t)=\left.w\tau\left(Z-\sum\nolimits _{z=1}^{Z}o_{z}(t)\right)\right/f_{{\rm L}},
\end{equation}
where ${f_{\rm{L}}}$ is the CPU computational frequency of the local VR device. Obviously, we have ${f_{\rm{L}}} < {f_{\rm{M}}}$.

Aside from the computational latency, the energy consumption for computation should also be taken into account, which is not that necessary for a traditional communication system. Let ${\eta _{\rm{M}}}$ and ${\eta _{\rm{L}}}$ denote the effective power switched capacitances of CPUs in MEC server and local VR device, respectively. The energy consumption for computing one cycle at the MEC server is ${\eta _{\rm{M}}}f_{\rm{M}}^2$ and similarly ${\eta _{\rm{L}}}f_{\rm{L}}^2$ for the local VR device \cite{8728029, Burd1996Processor}\footnote{Note that though the power model used in this work is simple, it has been widely adopted in the related works and its effectiveness has been validated  (c.f., e.g., \cite{8728029, Burd1996Processor}). We are fully aware of the fact that the power characteristics of different devices can vary wildly and standby/static power can account for a considerable portion of the total power. However, to avoid further complicating the problem under investigation, the establishment of a more realistic power model is left for our future work.}. Then the computation energy consumptions of the MEC server and the local VR device in each time slot $t$ can be respectively represented as

\begin{equation} \label{e5}
{E_{\rm{M}}}(t) = {\eta _{\rm{M}}}f_{\rm{M}}^2w\tau \sum\nolimits_{z = 1}^Z {{{\rm{o}}_z}(t)},
\end{equation}
\begin{equation} \label{e6}
{E_{\rm{L}}}(t) = {\eta _{\rm{L}}}f_{\rm{L}}^2w\tau \left( {Z - \sum\nolimits_{z = 1}^Z {{{\rm{o}}_z}(t)} } \right).
\end{equation}
Note that, in this work, we assume that the VR device has sufficient battery power to support long-time video service. That is, the limitation of the VR device's battery capacity is ignored. We will certainly consider this rather important constraint in our future work.

\subsection{Dynamic Caching Replacement Model}
\label{ss:DCRM}

In each time slot, once the local VR device and the MEC server receive new tiles for computation, the caching replacement should be simultaneously considered to make full use of the cache resources. We first consider the cache dynamic replacement at the local VR device. We assume that the cache size of the local VR device is $\tau {M_{\rm{L}}}$ (in bit), where ${M_{\rm{L}}}$ is an integer, and the cache state at the beginning of slot $t$ is represented as ${{\tilde {\cal F}}_{\rm{L}}}(t) = \{ F_1^{\rm{L}}(t), \cdots ,F_{{M_{\rm{L}}}}^{\rm{L}}(t)\} $. Let ${{\bf{c}}_{\rm{L}}}(t) = ({\bf{c}}_{\rm{L}}^ + (t),{\bf{c}}_{\rm{L}}^ - (t))$ denote the caching replacement action at the local VR device at slot $t$, where ${\bf{c}}_{\rm{L}}^ + (t) = [c_{{\rm{L}}1}^ + (t), \cdots ,c_{{\rm{L}}R(t)}^ + (t)]$ decides which tiles in ${{\cal F}_{\rm{L}}}(t)$ should be stored while ${\bf{c}}_{\rm{L}}^ - (t) = [c_{{\rm{L}}1}^ - (t), \cdots ,c_{{\rm{L}}{M_{\rm{L}}}}^ - (t)]$ decides which tiles in ${{\tilde {\cal F}}_{\rm{L}}}(t)$ should be deleted. $c_{{\rm{L}}r}^ + (t) = 1$ indicates that the tile ${F_{k_r^{\rm{L}}(t)}} \in {{\cal F}_{\rm{L}}}(t)$ should be stored; otherwise if $c_{{\rm{L}}r}^ + (t) = 0$, it should be outright discarded. Similarly, $c_{{\rm{L}}{m_{\rm{L}}}}^ - (t) = 1$ means that $F_{{m_{\rm{L}}}}^{\rm{L}}(t) \in {{\tilde {\cal F}}_{\rm{L}}}(t)$ should be deleted; otherwise if $c_{{\rm{L}}{m_{\rm{L}}}}^ - (t) = 0$, it should be preserved. Limited by the cache capacity of the local VR device, we have
\begin{equation} \label{e6.5}
\sum\nolimits_{r = 1}^{R(t)} {c_{{\rm{L}}r}^ + (t)}  = \sum\nolimits_{{m_{\rm{L}}} = 1}^{{M_{\rm{L}}}} {c_{{\rm{L}}{m_{\rm{L}}}}^ - (t)}.
\end{equation}
If $\{ {F_{k_r^{\rm{L}}(t)}}|c_{{\rm{L}}r}^ + (t) = 1,r = 1,2, \cdots ,R(t)\}  \cap \{ F_{{m_{\rm{L}}}}^{\rm{L}}(t)|c_{{\rm{L}}{m_{\rm{L}}}}^ - (t) = 1,{m_{\rm{L}}} = 1,2, \cdots ,{M_{\rm{L}}}\}  \ne \emptyset $, some tiles are included in the target sets of deletion and preservation at the same time. In this case, these tiles will remain unchanged.

We then consider the cache dynamic replacement at the MEC server. We suppose that its cache size is $\tau {M_{\rm{E}}}$ (in bit), where ${M_{\rm{E}}}$ is an integer and ${M_{\rm{E}}} > {M_{\rm{L}}}$. At the beginning of slot $t$, the original cache state is ${{\tilde {\cal F}}_{\rm{M}}}(t) = \{ F_1^{\rm{M}}(t), \cdots ,F_{{M_{\rm{E}}}}^{\rm{M}}(t)\}$. Similarly, we have ${{\bf{c}}_{\rm{M}}}(t) = ({\bf{c}}_{\rm{M}}^ + (t),{\bf{c}}_{\rm{M}}^ - (t))$, where ${\bf{c}}_{\rm{M}}^ + (t) = [c_{{\rm{M}}1}^ + (t), \cdots ,c_{{\rm{M}}S(t)}^ + (t)]$ and ${\bf{c}}_{\rm{M}}^ - (t) = [c_{{\rm{M}}1}^ - (t), \cdots ,c_{{\rm{M}}{M_{\rm{E}}}}^ - (t)]$. $c_{Ms}^ + (t) = 1$ and $c_{M{m_{\rm{E}}}}^ - (t) = 1$ respectively means that the tile ${F_{k_s^M(t)}} \in {{\cal F}_M}(t)$ should be preserved and the tile $F_{{m_{\rm{E}}}}^{\rm{M}}(t) \in {{\tilde {\cal F}}_{\rm{M}}}(t)$ should be deleted. We can also obtain the following cache capacity constraint
\begin{equation} \label{e7}
\sum\nolimits_{s = 1}^{S(t)} {c_{{\rm{M}}s}^ + (t)}  = \sum\nolimits_{{m_{\rm{E}}} = 1}^{{M_{\rm{E}}}} {c_{{\rm{M}}{m_{\rm{E}}}}^ - (t)}.
\end{equation}

Supported by the caching component, this process can be completed before the computation such that no additional latency is induced. In addition, as the energy consumption caused by accessing files is significantly lower than that of the computation, we neglect the former in the modeling.

\subsection{Wireless Transmission Model}

There are three types of transmission links in the MEC-assisted VR service system, i.e., the uplink, the downlink, and the backhual link. In the uplink, only the viewpoint request and the cache state information are uploaded to the MEC server. As such, the size of the transmitted data is negligible compared to the size of the 2D tiles and the 3D FoV files. We therefore ignore the latency cost in the uplink.

Regarding the downlink transmission, given the bandwidth $B$ and the transmit power ${P_B}$, the wireless downlink rate is derived as
\begin{equation} \label{e8}
{R_{\rm{wl}}}(t) = B \cdot {\log _2}(1 + \ell (t)),
\end{equation}
where $\ell (t) = {{{P_B} \cdot h(t)}}/{{{\sigma ^2}}}$ is the signal-to-interference-plus-noise ratio (SINR) in a slot $t$. ${{\sigma ^2}}$ denotes the variance of the Gaussian noise and $h(t) = g(t) \cdot {d^{ - \alpha }}$ represents the path loss between the SBS and the VR user with $g(t)$ being the Rayleigh fading parameter and assumed to be constant within one slot duration. $d$ represents the distance between the user and the SBS, and $\alpha$ is the path loss exponent. Let ${{\bf{C}}^{\rm{L}}}(t) = \{C_i^{\rm{L}}(t)\}_{i = 1}^N$ be the vector representation of the cache state in the local VR device at the beginning of time slot $t$, where each entry corresponds to an index in ${\cal F}$. $C_i^{\rm{L}}(t) = 1$ means that the tile ${F_{\rm{i}}}$ is cached at the local, and $\sum_{i = 1}^N{C_i^{\rm{L}}(t)} \le {M_{\rm{L}}}$ is the cache capacity constraint. Likewise, we have ${{\bf{C}}^{\rm{M}}}(t) = \{C_i^{\rm{M}}(t)\}_{i = 1}^N$ and $\sum_{i = 1}^N {C_i^{\rm{M}}(t)} \le {M_{\rm{E}}}$ for the MEC server. In a time slot $t$, the size of 2D tiles transmitted through the downlink, which are demanded but absent in the local VR device, can be denoted as
\begin{equation}\label{e9}
\begin{array}{l}
D_{\rm{2D}}^{\rm{down}}(t) =\displaystyle \tau  \cdot \sum\nolimits_{i = 1}^Z {{{\bf{1}}_{({o_i}(t) = 0\& C_i^{\rm{L}}(t) = 0)}}}\\
\qquad\displaystyle= \tau  \cdot \sum\nolimits_{i = 1}^Z {{{\bf{1}}_{({o_i}(t) = 0\& C_i^{\rm{L}}(t) = 0\& C_i^{\rm{M}}(t) = 1)}}}\\
\qquad\displaystyle+\tau  \cdot \sum\limits_{i = 1}^Z {{{\bf{1}}_{({o_i}(t) = 0\& C_i^{\rm{L}}(t) = 0\& C_i^{\rm{M}}(t) = 0)}}}\\
\qquad\displaystyle= D_{\rm{mec}}^{\rm{down}}(t) + D_{\rm{cloud}}^{\rm{down}}(t),
\end{array}
\end{equation}
where the first term on the righthand side is the size of the tiles directly accessed from the MEC server, and the second term is the size of the tiles accessed from the cloud because neither does the MEC server have a copy of the requested tiles. The size of the 3D file transmitted by the downlink, which is the computed output of the MEC server, is given by
\begin{equation} \label{e10}
D_{\rm{3D}}^{\rm{down}}(t) = \tau  \cdot \varphi  \cdot \sum\nolimits_{i = 1}^Z {{{\rm{o}}_z}(t)}.
\end{equation}
Thus, the latency cost of the downlink transmission is
\begin{equation} \label{e11}
T^{{\rm down}}(t)=\left.\left(D_{{\rm 2D}}^{{\rm down}}(t)+D_{{\rm 3D}}^{{\rm down}}(t)\right)\right/R_{{\rm wl}}(t).
\end{equation}

We now consider the backhual link in the wireless transmission model. Given the bandwidth ${{R_{\rm{bh}}}}$, the transmission latency on the backhual link can be expressed as
\begin{equation} \label{e12}
T^{{\rm back}}(t)=\left.\left(D_{{\rm mec}}^{{\rm back}}(t)+D_{{\rm local}}^{{\rm back}}(t)\right)\right/R_{{\rm bh}},
\end{equation}
where $D_{\rm{mec}}^{\rm{back}}(t){\rm{ = }}\tau  \cdot \sum_{i = 1}^Z {{{\bf{1}}_{({o_i}(t) = 1\& C_i^{\rm{M}}(t) = 0)}}}$ is the size of the tiles absent in the MEC server but demanded by MEC server's computation, while $D_{\rm{local}}^{\rm{back}}(t) = D_{\rm{cloud}}^{\rm{down}}(t)$ denotes the size of the tiles absent both in the local VR device and the MEC server but computed in the local VR device. It is noteworthy that the overall latency is not simply a sum of all the latencies introduced in this subsection. We will elaborated this issue in following subsection.

In the wireless transmission model of this work, the system energy is mainly used for the transmission from the SBS to the local VR device. As such, the system energy consumption can be approximated as
\begin{equation} \label{e13}
{E_{\rm{T}}}(t) = {P_B} \cdot {T^{\rm{down}}}(t).
\end{equation}

\subsection{Utility Function Model}

We jointly consider both the latency and energy costs in the utility evaluation.

We first look into the latency cost for each task, which is also defined as the duration of each time slot. As deterministic multi-tile offloading is implemented in the proposed scheme, the task is decomposed into two subtasks and processed in parallel at the both the MEC server and the local VR device. For the MEC server, we have
\begin{equation} \label{e14}
{T^{\rm{M}}}(t) = \frac{{D_{\rm{mec}}^{\rm{back}}(t)}}{{{R_{\rm{bh}}}}} + \frac{{D_{\rm{3D}}^{\rm{down}}(t)}}{{{R_{\rm{wl}}}(t)}} + T_{\rm{com}}^{\rm{M}}(t),
\end{equation}
where the first term at the righthand side is the transmission time of the 2D tiles from the backhual link. These tiles are used for the computation at the MEC server in slot $t$, which are demanded by the MEC server but cannot be found in its own cache. The second term is the transmission time of MEC server's computation output content, which is a portion of the 3D FoV requested by the user in slot $t$. The rightmost term is the computation time of the MEC server. For the local VR device, we define its corresponding latency cost as
\begin{equation} \label{e15}
{T^{\rm{L}}}(t){\rm{ = }}\max \left\{ \frac{{D_{\rm{local}}^{\rm{back}}(t)}}{{{R_{\rm{bh}}}}},\frac{{D_{\rm{mec}}^{\rm{down}}(t)}}{{{R_{\rm{wl}}}(t)}}\right\}  + \frac{{D_{\rm{local}}^{\rm{back}}(t)}}{{{R_{\rm{wl}}}(t)}} + T_{\rm{com}}^{\rm{L}}(t),
\end{equation}
where $\frac{{D_{\rm{local}}^{\rm{back}}(t)}}{{{R_{\rm{bh}}}}}$ is the transmission time of the 2D tiles from the cloud to the SBS by backhual link, which are demanded by the local VR device, but absent both in the local and the MEC server. $\frac{{D_{\rm{mec}}^{\rm{down}}(t)}}{{{R_{\rm{wl}}}(t)}}$ is the transmission time of the 2D tiles that cannot be found in the local cache but are stored in the MEC server. The latency is determined by the maximum of these two transmission durations due to the parallel transmission in the downlink and backhual link. After receiving the 2D tiles from the cloud, the SBS forwards them to the local VR device by downlink. The induced latency is denoted as $\frac{{D_{\rm{local}}^{\rm{back}}(t)}}{{{R_{\rm{wl}}}(t)}}$. The rightmost term represents the computation time of the local VR device. Because of the parallel processing of the two devices, the overall service latency can be derived as
\begin{equation} \label{e16}
{T^{\rm{total}}}(t) = \max \left\{ {T^{\rm{M}}}(t),{T^{\rm{L}}}(t)\right\}.
\end{equation}
Next, we define the energy cost generated by the computation and the transmission, i.e.,
\begin{equation} \label{e17}
{E^{\rm{total}}}(t) = {E_{\rm{M}}}(t) + {E_{\rm{L}}}(t) + {E_{\rm{T}}}(t).
\end{equation}

To find a latency-energy tradeoff represents an important research direction in wireless VR delivery. Many related works suggest the weighted-sum method \cite{8677285, Trinh18, Zhang19}, which is intuitive and quite effective. We follow this path and define the cost in each time slot as
\begin{equation} \label{e18}
Y(t) = \omega {T^{\rm{total}}}(t) + (1 - \omega ){E^{\rm{total}}}(t),
\end{equation}
where $\omega  \in [0,1]$ is the weight that indicates the relative importance of the ${T^{\rm{total}}}(t)$ and the ${E^{\rm{total}}}(t)$. However, in providing VR video service, we pay more attention to the long-term experience during a continuous period of time rather than the instantaneous experience. In this case, suppose that $\Upsilon$ time slots are examined to evaluate the overall utility of the system. Then, the cumulative utility over time is given by
\begin{equation} \label{e19}
U =  - \sum\nolimits_{t = 1}^\Upsilon  {Y(t)}.
\end{equation}

\section{PROBLEM FORMULATION}
\label{sec3}

\subsection{Problem Formulation}

As mentioned earlier, we concerned more about the long-term performance over a continuous period of time rather than the instantaneous performance in wireless VR service. Therefore, the designed hybrid policy aims at minimizing the cumulative cost of the VR service system. As such, the underlying optimization problem is formulated as
\begin{subequations}\label{e24}
    \begin{align}
    \mathop {\min }\limits_{({\cal O},{{\cal C}_{\rm{L}}},{{\cal C}_{\rm{M}}})} \quad
    &\mathop {\lim }\limits_{\Upsilon  \to \infty } \sum\nolimits_{k = 0}^\Upsilon  {{\mathbb{E}}\left[ {{\chi ^k}Y(t+k)} \right]},\label{e24a}\\
    {\rm{s.t.}}\quad \quad \quad
    &\sum\nolimits_{i = 1}^N {C_i^{\rm{L}}(t)}  \le  {M_{\rm{L}}},\label{e24b}\\
    &\sum\nolimits_{i = 1}^N {C_i^{\rm{M}}(t)}  \le {M_{\rm{E}}},\label{e24c}\\
    &\sum\nolimits_{r = 1}^{R(t)} {c_{{\rm{L}}r}^ + (t)}  = \sum\nolimits_{{m_{\rm{L}}} = 1}^{{M_{\rm{L}}}} {c_{{\rm{L}}{m_{\rm{L}}}}^ - (t)},\label{e24d}\\
    &\sum\nolimits_{s = 1}^{S(t)} {c_{{\rm{M}}s}^ + (t)}  = \sum\nolimits_{{m_{\rm{E}}} = 1}^{{M_{\rm{E}}}} {c_{{\rm{M}}{m_{\rm{E}}}}^ - (t)},\label{e24e}\\
    &{o_z}(t) \in \{ 0,1\} ,\;\forall z \in {\cal Z},\label{e24f}\\
    &c_{{\rm{L}}r}^ + (t),c_{{\rm{L}}{m_{\rm{L}}}}^ - (t) \in \{ 0,1\},\nonumber \\
    &\qquad\quad(\forall r \in {{\cal U}_{\rm{L}}}(t),\forall {m_{\rm{L}}} \in {{\cal M}_{\rm{L}}}),\label{e24g}\\
    &c_{{\rm{M}}s}^ + (t),c_{{\rm{M}}{m_{\rm{E}}}}^ - (t) \in \{ 0,1\},\nonumber \\
    &\qquad\quad(\forall s \in {{\cal U}_{\rm{M}}}(t),\forall {m_{\rm{E}}} \in {{\cal M}_{\rm{M}}}),\label{e24h}
    \end{align}
\end{subequations}
where the optimization parameter sets ${\cal O} = \{ {\bf{o}}(t)|t = 0,1,2,3, \cdots \}$, ${{\cal C}_{\rm{L}}} = {\rm{\{ }}{{\bf{c}}_{\rm{L}}}(t)|t = 0,1,2,3, \cdots {\rm{\} }}$, and ${{\cal C}_{\rm{M}}} = {\rm{\{ }}{{\bf{c}}_{\rm{M}}}(t){\rm{|}}t = 0,1,2,3, \cdots {\rm{\} }}$ represent the collections of actions in each time slot $t$. ${{{\bf{c}}_{\rm{L}}}(t)}$ contains two vectors, namely ${\bf{c}}_{\rm{L}}^ + (t)$ and ${\bf{c}}_{\rm{L}}^ - (t)$, which have been defined in Section \ref{ss:DCRM}. Likewise, ${{\bf{c}}_{\rm{M}}}(t) = ({\bf{c}}_{\rm{M}}^ + (t),{\bf{c}}_{\rm{M}}^ - (t))$. ${Y(t)}$ denotes the joint latency/energy cost at any time t. $\chi  \in (0,1)$ is the discount factor. The expectation is taken with respect to the measure included by the decision variables as well as the system state. ${{\cal U}_{\rm{L}}}(t) = \{ 1,2, \cdots ,R(t)\}$ and ${{\cal U}_{\rm{M}}}(t) = \{ 1,2, \cdots ,S(t)\}$ represent the time-varying index sets of the  computed tiles sets ${{\cal F}_{\rm{L}}}(t)$ and ${{\cal F}_{\rm{M}}}(t)$, respectively. ${{\cal M}_{\rm{L}}} = \{ 1,2, \cdots ,{M_{\rm{L}}}\}$ and ${{\cal M}_{\rm{M}}} = \{ 1,2, \cdots ,{M_{\rm{E}}}\}$ are the index sets of ${{\tilde {\cal F}}_{\rm{L}}}(t)$ and ${{\tilde {\cal F}}_{\rm{M}}}(t)$, respectively. Additionally, ${\cal Z} = \{ 1,2, \cdots ,Z\}$. Here we assume that the unknown viewpoint request distribution of the user varies over time. Note that, in a realistic environment, the request distribution may remain unchanged during a period. This is equivalent to a special case of our model where the distribution parameters transit to themselves during the period.

The constraints in \eqref{e24b} and \eqref{e24c} indicate that the number of cached files in the local VR device and the MEC server are both limited by the cache capacity. To maximize the utilization of the caching resources, the caches in both of the devices should be fully filled. On the other hand, the constraints in \eqref{e24d} and \eqref{e24e} respectively ensures a balance in the sizes of the cached files at the local VR device and the MEC device after the caching replacement, in order to keep the caches full but not overflowed. The constraint in \eqref{e24f} implies that each tile of the computation input ${F_{{k_z}(t)}}$ can be computed at either the local VR device or the MEC server. \eqref{e24g} indicates that the caching replacement in each time slot in the local VR device is limited by ${{\cal F}_{\rm{L}}}(t)$ and ${{\tilde {\cal F}}_{\rm{L}}}(t)$, which guarantees that the caching replacement in the local VR device does not bring any additional resource consumption. \eqref{e24h} plays a similar role for the caching operation in the MEC server.
\begin{figure*}[!pbth]
\normalsize
\newcounter{MYtempeqncnt}
\begin{equation} \label{theorem1proof}
\begin{array}{l}
V(s(t)) = \mathop {\lim }\limits_{\Upsilon  \to \infty } \sum\limits_{k = 0}^\Upsilon  {{{\mathbb{E}}_\pi }\left[ {{\chi ^k}Y(t + k){|_{s(t)}}} \right]}
= {{\mathbb{E}}_\pi }\left[ {\sum\limits_{k = 0}^\infty  {{\chi ^k}Y(t + k){|_{s(t)}}} } \right]\\
\quad\quad\quad = {{\mathbb{E}}_\pi }\left[ {Y(t){|_{s(t)}} + \sum\limits_{k = 0}^\infty  {{\chi ^k}Y(t + k){|_{s(t)}}} } \right]
= {{\mathbb{E}}_\pi }\left[ {Y(t){|_{s(t)}}} \right] + {{\mathbb{E}}_\pi }\left[ {\sum\limits_{k = 0}^\infty  {{\chi ^k}Y(t + k + 1){|_{s(t)}}} } \right]\\
\quad\quad\quad = \sum\limits_a {\pi (s(t),a)\sum\limits_{s(t{\rm{ + }}1)} {P_{s(t) \to s(t + 1)}^aR_{s(t) \to s(t)}^a} }  +
\chi \sum\limits_a {\pi (s(t),a)\sum\limits_{s(t + 1)} {P_{s(t) \to s(t)}^a\mathop {\lim }\limits_{\Upsilon  \to \infty } \sum\limits_{k = 0}^\Upsilon  {{{\mathbb{E}}_\pi }\left[{\chi ^k}Y(t + 1{\rm{ + }}k){|_{s(t + 1)}}\right]} } } \\
\quad\quad\quad = \sum\limits_a {\pi (s(t),a)\sum\limits_{s(t + 1)} {P_{s(t) \to s(t + 1)}^a\left[R_{s(t) \to s(t + 1)}^a + \chi V(s(t + 1))\right]} }
\end{array}
\end{equation}
\vspace{-0.36cm}
\hrulefill
\end{figure*}
The following facts and technical challenges should be noted:
\begin{itemize}
  \item The solution of the problem under investigation is a dynamic strategy over time rather than a transient one.
  \item The system states and actions conform to chain property over time.
  \item The dimensions of the two actions ${{\bf{c}}_{\rm{L}}}(t)$ and ${{{\bf{c}}_{\rm{M}}}(t)}$ are both time-varying.
  \item The problem \eqref{e24} corresponds to an infinite-horizon-cost MDP problem \cite{8677285} and \emph{a priori} knowledge on state transition probabilities is unavailable.
\end{itemize}
To overcome the above fourfold technical challenges, we exploit deep reinforcement learning in this work instead of traditional optimization techniques such as dynamic programming. The proposed scheme offers inherent advantages in addressing the infinite-horizon-cost MDP problems, complicated and time-varying system states, continuous strategy development, and etc.
\begin{theorem}\label{theorem1}
Without any knowledge of state transition probabilities $P_{s(t) \to s(t + 1)}^a = P\{ s(t + 1)|s(t),a\}$, the problem \eqref{e24} is an infinite-horizon-cost MDP problem.
\end{theorem}
\begin{IEEEproof}
Let $\pi (s(t),a)$ denote the probability of selecting action $a$ under strategy $\pi $ when the system state is $s(t)$ . The objective function in \eqref{e24} is denoted as $V(s(t)) = \mathop {\lim }\limits_{\Upsilon  \to \infty } \sum\nolimits_{k = 0}^\Upsilon  {{_\pi }[{\chi ^k}Y(t}$ $+ k){|_{s(t)}}]$. The derivation is given in (\ref{theorem1proof}), where ${R_{s(t) \to s(t + 1)}^a}$ is the reward feedback when the system state transits from $s(t)$ to $s(t+1)$ under action $a$.
We observe from the above derivation that, the analytical solution of $V(s(t))$ is unobtainable since $P_{s(t) \to s(t + 1)}^a$ is unknown and the term $V(s(t + 1))$ is recursive in a given time slot $t$. As such, we cannot get the exact value of the objective function in problem \eqref{e24} even if the decision is executed. Therefore, the problem \eqref{e24} is an infinite-horizon-cost MDP problem.
\end{IEEEproof}

\subsection{MDP Description}

To address the above difficulty, we firstly convert our problem into a MDP which consists of four components, i.e, state space, action space, state transition probabilities, and reward. The MDP description of the optimization problem is denoted as $<{\cal S},{\cal A},{\bf{P}},R>$. Note that in this work, we assume that the state transition probabilities of the Markov model is unknown and the state is partly observable.

\subsubsection{State}

At the beginning of each slot, the local VR device will upload its current cache index information ${{\tilde {\cal F}}_{\rm{L}}}(t)$ along with its viewpoint request. At the same time, the MEC server also awares of its own cache index information ${{\tilde {\cal F}}_{\rm{M}}}(t)$. We assume that the path loss $h(t)$ between the SBS and the VR user remains unchanged in a slot duration and can be sensed by the SBS in realtime. As already defined in Section \ref{sec2}, ${\bf{\hat p}}(t + 1)$ is used as a prediction of the unknown view popularity, whose implementation details will be explained in Section \ref{sec4}. Then, the current system state is modelled by the following combination of variables:
\begin{equation} \label{e20}
{\bf{s}}(t) = \left\{ \left({{\tilde {\cal F}}_{\rm{L}}}(t),{{\tilde {\cal F}}_{\rm{M}}}(t),{\bf{\hat p}}(t + 1),h(t)\right)\right\}.
\end{equation}

\subsubsection{Action}

The set of possible actions ${\bf{a}}(t) \in {\cal A}$ is composed of the set of deterministic offloading action ${\bf{o}}(t)$, and the cache replacement policies ${{\bf{c}}_{\rm{L}}}(t)$ and  ${{\bf{c}}_{\rm{M}}}(t)$ at the local VR device and the MEC server. Let $\pi$ be the policy and it will output actions in any state ${\bf{s}}(t) \in {\cal S}$. Hence, the actions under state ${\bf{s}}(t)$ in slot $t$ is given by
\begin{equation} \label{e21}
{\bf{a}}(t) = \pi ({\bf{s}}(t)) = \left\{ \left({\bf{o}}(t),{{\bf{c}}_{\rm{L}}}(t),{{\bf{c}}_{\rm{M}}}(t)\right)\right\}.
\end{equation}

\subsubsection{State Transition}

$\forall {\bf{s}}(t),{\bf{s}}(t + 1) \in {\cal S}$ and $\forall {\bf{a}}(t) \in {\cal A}$, the state transition probability $P_{{\bf{s}}(t) \to {\bf{s}}(t)}^{{\bf{a}}(t+1)}$ describes the probability that a system transits from the state ${{\bf{s}}(t)}$ to the state ${{\bf{s}}(t{\rm{ + }}1)}$ under the action ${{\bf{a}}(t)}$. As for the solution of traditional optimization approach or the dynamic programming, the state transition probability is an essential component which however is often not available in the real world due to the complicated and changeable environment. Here we just explain the meaning of the state transition probability in Markov problem. In our problem, on account of the view popularity's dynamics and the channel condition's randomness, we assume that the state transition probability is unknown.

\subsubsection{Reward Function}

The reward function assigns each perceived state to a value associated with an explicit goal in the reinforcement learning problem. For an MDP, when an action is taken under a state, the state will transfer to another state and the environment will return an instantaneous reward as a feedback immediately, which is derived as a negative of the cost $Y(t)$ in our problem, i.e.,
\begin{equation} \label{e22}
r(t) =  - Y(t).
\end{equation}
On this basis, we define the cumulative reward as
\begin{equation}
R =   \sum\nolimits_{t = 1}^\Upsilon  {{\chi ^t}r(t)},
\end{equation}
where $\chi  \in (0,1)$ is the discount factor. The discounted definition is common adopted in MDP problems and  is similar to the definition of value function in reinforcement learning.

\section{METHODOLOGY}\label{sec4}

\begin{figure*}[!htpb]
\centering
\includegraphics[width=0.34\textwidth,angle=270]{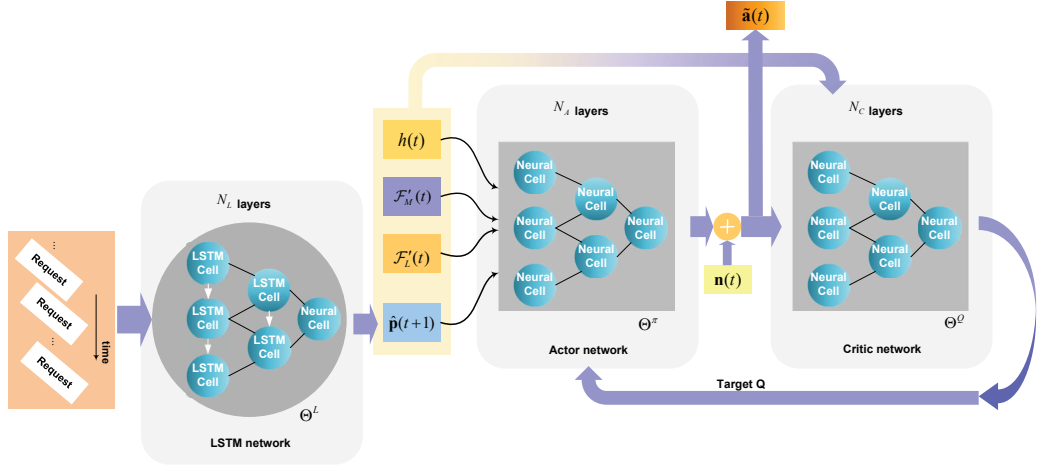}
\caption{Network architecture of the LSTM-DDPG algorithm.}
\label{fig5}
\vspace{-1em}
\end{figure*}

In this section, we first develop an LSTM neural network to predict the unknown and time-varying viewpoint popularity. Then, we integrate the LSTM network \cite{Suts14, Su18} into the DDPG algorithm \cite{Silver2014Deterministic, Lillicrap2015Continuous} to iteratively learn the optimal hybrid policy for cooperative multi-tile offloading and caching replacement such that the system cumulative cost is minimized. The network architecture is illustrated in Fig. \ref{fig5}.

\subsection{Popularity Prediction Based on LSTM}

It is more reasonable and realistic to assume that the \emph{a priori} knowledge on the viewpoint popularity cannot be acquired. As mentioned above, in this paper the dynamic popularity ${{\bf{p}}^{{\gamma _t}}}(t)$ is modeled as a Zipf distribution whose parameters follow a Markov stochastic chain and transfer in a certain parameter space. However, neither the parameter set nor the transition probability is accessible for the system. The only information the system knows exactly is the user's requests under the popularity in each slot, which is written as ${V_k}(t){|_{{{\bf{p}}^{{\gamma _t}}}(t)}}$ and abbreviated as ${V_k}(t)$. Taking full advantage of the request information, we design a multilayer LSTM neural network with three layers including two LSTM layers and one dense layer to predict the viewpoints popularity.

At the beginning of time slot $t$, the current viewpoint request ${V_k}(t)$ is added to the recorder and meanwhile the oldest request ${V_k}(t - {T_{\rm{r}}})$ is removed in order to keep the length of the recorder constant. The requests vector ${\bf{R}}(t)$ is then obtained by stacking a series of ${T_r}$ requests in successive time slots, i.e., as ${\bf{R}}(t) = [{V_k}(t - {T_{\rm{r}}} + 1), \cdots ,{V_k}(t)]$. Subsequently, these requests are fed to the LSTM neural network in sequence. Note that, in the LSTM neural network, the input in each slot contains not only the external input ${V_k}(t - 1)$ but also the internal input, which consists of the output and the memory state from the previous slot. After the feeding of ${T_r}$ successive requests, the estimated popularity of the viewpoints request is outputted in the next time slot $t+1$. Referring to the definition in \eqref{e2}, we have
\begin{equation} \label{e25}
{\bf{\hat p}}(t + 1) = {f_{{\Theta ^{\rm{L}}}}}\left({\bf{R}}(t)\right),
\end{equation}
where ${\Theta ^{\rm{L}}}$ is the collection of the trainable parameters and ${\bf{\hat p}}(t + 1)$ is the predicted popularity. The training set of the LSTM network can be denotes as $\{ ({{\bf{x}}^i},{{\bf{y}}^i})|{{\bf{x}}^i} = {\bf{R}}({t_i}),{{\bf{y}}^i} = {{\bf{p}}^{{\gamma _t}}}({t_i} + 1)\}$, which is acquired from the historical experience. The time slot set corresponding to the training set is represented as $\{ {t_1}, \cdots ,{t_{\cal L}}\}$, and the batch average of the MSE loss function is adopted to yield a more stable convergence. Thus, for any batch set $\{ {t_{{i_1}}}, \cdots ,{t_{{i_X}}}\}  \subseteq \{ {t_1}, \cdots ,{t_{\cal L}}\}$ where ${i_x} \in \{ 1, \cdots ,{\cal L}\} $, we have
\begin{equation} \label{e26}
L({\Theta ^{\rm{L}}}) = \frac{1}{X}\sum\nolimits_{x = 1}^X {{{\left| {{\bf{\hat p}}({t_{{i_x}}} + 1) - {{\bf{p}}^{{\gamma _t}}}({t_{{i_x}}} + 1)} \right|}^2}} ,
\end{equation}
where $X$ denotes the batch size. The primary goal of the LSTM is to minimize $L({\Theta ^{\rm{L}}})$ and find the optimal ${\Theta ^{\rm{L}}}$ via gradient descent. After the convergence of training, the LSTM networks module can be embedded into the DDPG algorithm for subsequent learning. The future popularity prediction affects the caching replacement policy and affects the offloading action in turn. Generally, high prediction accuracy yields high system performance.

\subsection{Learning Hybrid Policy Using LSTM-DDPG}

The optimal policy ${\pi ^ * }$ determines which action should be selected from the action space at any time $t$ to maximize the cumulative reward function. For traditional dynamic programming algorithms, such policy is derived recursively from the Bellman's Equation and its solution can be written as
\begin{equation} \label{e27}
\begin{array}{l}
\pi^{*}=\\
\mathop{\arg\;\max}\limits _{{\bf a}(t)\in A}\sum\nolimits _{{\bf s}(t{\rm +}1)\in S}P_{{\bf s}(t)\to{\bf s}(t+1)}^{{\bf a}(t)}\left(r\left(t\right)+\chi V^{\pi}\left({\bf s}\left(t+1\right)\right)\right),
\end{array}
\end{equation}
where ${{V^\pi }\left( {{\bf{s}}\left( {t + 1} \right)} \right)}$ is the value function which indicates the cumulative reward under policy $\pi$ at state ${{\bf{s}}(t + 1)}$. To obtain the optimal ${\pi ^ * }$ in such a manner, the transition probability ${P_{{\bf{s}}(t) \to {\bf{s}}(t + 1)}^{{\bf{a}}(t)}}$ is essential. As a component of the system state, although the future popularity of the viewpoint request can be predicted, the system still has no idea about the distributed parameter space, or even the number of different states. Therefore, the formulated Markov domain lacks the state transition mapping as addressed in Remark 1. As such, the traditional dynamic programming algorithms are inapplicable to the problem under investigation in this work.

In this paper we focus on the deep reinforcement learning approach for the formulated problem \eqref{e24}. Conventional learning approaches such as Q-learning usually build a Q-table to record the Q-value whose size increase exponentially along with the size of state/action spaces. The deep Q-network (DQN) algorithm \cite{Volodymyr2015Human, Mnih13} replaces the Q-table with a neural network and can be used for the situations with large/continuous state spaces. However, it still needs to select the actions according to the maximal Q-value based on the output of the neural network, which makes it computational intensive for the situations with large/continuous action space. In our problem, the state space ${\cal S}$ is a mixed space with the caching state belonging to a discrete space while the popularity and channel states belonging to a continuous spaces. The action space ${\cal A}$ is extremely large because the multi-tile has ${2^{2Z + {M_{\rm{L}}} + {M_{\rm{E}}}}}$ possible values in a single time slot. Thus, neither can the DQN algorithm be applied. The DDPG algorithm directly parameterizes the policy ${\pi _{{\Theta _2}}}$ rather than the Q-value function as in DQN or Q-learning, which enables the LSTM-DDPG algorithm to handle problems with large and continuous action space. Moreover, the user's request distributions, as one important component of the state, will change as time elapses and is totally unknown to the agent. We embed the LSTM neural network into the DDPG algorithm to counteract with this limitation and improve the system performance and convergence speed simultaneously.

The structure of the proposed algorithm to solve problem (21) is shown in Fig. \ref{fig5}. The deep neural network is an important building block of the entire network structure. ${{\cal F}_{{V_k}(t)}}$ is vectorized as ${\rm {vec}}({\cal F}_{{V_k}(t)}) = [x_k^u(t)]_{u = 1}^N$ before being fed into the neural network, where ${\rm{x}}_k^u{\rm{(t)}} = 1$ indicates that the tile ${F_u}$ is in the viewpoint ${V_k}(t)$; and ${\rm{x}}_k^u{\rm{(t)}} = 0$ otherwise. Then, we input the recorded request vector ${\bf{R}}(t) = \left[{\rm{vec}}{\rm{(}}{{\cal F}_{{V_k}(t - {T_{\rm{r}}} + 1)}}{\rm{)}}, \cdots ,{\rm{vec}}{\rm{(}}{{\cal F}_{{V_k}(t)}}{\rm{)}}\right]$ into the ${N_{\rm{L}}}$-layer LSTM neural network and obtain ${\bf{\hat p}}(t + 1)$. Likewise, we vectorize the ${{\tilde {\cal F}}_{\rm{L}}}(t)$ and  ${{\tilde {\cal F}}_{\rm{M}}}(t)$ as ${\rm{vec}}({{\tilde {\cal F}}_{\rm{L}}}(t)) = \left[x_{\rm{L}}^i\right]_{i = 1}^N$ and ${\rm{vec}}({{\tilde {\cal F}}_{\rm{M}}}(t)) = \left[x_{\rm{M}}^j\right]_{j = 1}^N$, respectively. Subsequently, the vectorial state ${\bf{s}}(t) = ({\rm{vec}}({{\tilde {\cal F}}_{\rm{L}}}(t)),{\rm{vec}}({{\tilde {\cal F}}_{\rm{M}}}(t)),{\bf{p'}}(t + 1),h(t))$ is fed into the actor network, which is also a neural network with ${N_{\rm{A}}}$ dense layers. The actor network's output consists of three elements, i.e., ${\bf{o}}(t)$, ${{\bf{c}}_{\rm{L}}}(t)$, and ${{\bf{c}}_{\rm{M}}}(t)$, which constitute ${\bf{a}}(t)$.

\begin{remark}\label{remark2}
In practice, the actor network outputs a continuous-valued vector and we need to convert the numeric values to the binary values of 0/1.

In this paper, we further add five neural cells to the output of the actor network to generate the threshold output for ${\bf{o}}(t),{\bf{c}}_{\rm{L}}^ + (t),{\bf{c}}_{\rm{L}}^ - (t),{\bf{c}}_{\rm{M}}^ + (t)$, and ${\bf{c}}_{\rm{M}}^ - (t)$ rather than manually setting thresholds. Therefore, the actual output of the actor network can be denoted as ${\bf{{\mathord{\buildrel{\lower3pt\hbox{$\scriptscriptstyle\smile$}}\over a} }}}(t){\rm{ = }}\left[{\bf{a}}(t),{\varepsilon _{\rm{o}}},\varepsilon _L^ + ,\varepsilon _L^ - ,\varepsilon _M^ + ,\varepsilon _M^ - \right]$. Nevertheless, we keep on using the notation ${\bf{a}}(t)$ in the following discussions for better illustration.
\end{remark}

In order for the agent to fully explore the environment, exploration-exploitation method is adopted. Different from the $\varepsilon$-greedy exploration \cite{Sutton1998Reinforcement} which is effective for the small or discrete action space. In this work, we balance the exploration and the exploitation by adding a gaussian noise vector on the policy output, i.e.,
\begin{equation} \label{e27.5}
{\bf{\tilde a}}(t) = {\bf{a}}(t) + {\bf{n}}(t){{\rm{|}}_{{n_i}(t) \sim {\cal N}(0,\sigma _n^2)}}.
\end{equation}
Then the action ${\bf{\tilde a}}(t)$ will be immediately sent to the critic neural network which contains ${N_{\rm{C}}}$ dense layers together with the vectorial state ${\bf{s}}(t)$. Consequently, the critic network outputs the target-Q $Q\left({\bf{s}}(t + 1),{\bf{\tilde a}}(t + 1)|{\Theta ^{\rm{Q}}}\right)$ which is a step forward for estimating the Q-value. ${\Theta ^{\rm{Q}}}$ is the collection of trainable parameters of the critic network. After a linear transformation, the output of the critic network is sent to the actor network and contribute to the actor's loss function. In this way, the LSTM neural network, the actor network, and the critic network are connected and work collaboratively.

In reinforcement learning, there are two important function called state value function denoted as $V(s)$ and state-action value function (Q function) denoted as $Q(s,a)$. $V(s)$ represents the cumulative reward starting from state $s$ under a certain policy $\pi$. In this paper, we adopt the discount method and the value function is thus defined as
\begin{equation} \label{e28}
V_\gamma ^\pi (s) = {E_\pi }\left[\sum\nolimits_{t = 0}^{ + \infty } {{\chi ^t}r(t + 1)|s(0) = s}\right],
\end{equation}
where $r(t)$ is the instantaneous reward. Besides, after the agent taking an action $a$ under the initial state $s$, the cumulative reward under a certain policy $\pi$ can be given by another function $Q(s,a)$, defined as
\begin{equation} \label{e29}
Q_\gamma ^\pi (s,a) = {E_\pi }\left[\sum\nolimits_{t = 0}^{ + \infty } {{\chi ^t}r(t + 1)|s(0) = s} ,a(0) = a\right].
\end{equation}
These two function are usually used for policy evaluation. We can see from their definitions that, the specific expectation cannot be obtained if the state transition probability is unknown. Actually, in reinforcement learning, they are given by sampling estimates.

Let $\pi (s|{\Theta ^\pi })$ denote the output of the actor network which is a policy on parameters ${\Theta ^\pi }$ and let $Q(s,a|{\Theta ^{\rm{Q}}})$ be the critic network's output on parameters ${\Theta ^{\rm{Q}}}$. In practical training, these two networks are called online networks. Correspondingly, there are two clone networks respectively called target actor network and target critic network, whose parameters are copied from their online counterparts every a few steps. We denote their output as $\pi'(s|{\Theta ^{\pi'}})$ and $Q'(s,a|{\Theta ^{{\rm{Q}}'}})$, respectively. These target networks are employed for a stabler convergence.

During the training phase, experience replay is adopted. We randomly take ${X_2}$ samples $\{ ({{\bf{s}}_i},{{{\bf{\tilde a}}}_i},{r_i},{{{\bf{s'}}}_i})\} _{i = 1}^{{X_2}}$ as a mini-batch from the replay memory buffer $\Omega$, where ${{{\bf{s'}}}_i}$ is the next state following state ${{\bf{s}}_i}$. Then, we train the online critic network to let the $Q(s,a|{\Theta ^{\rm{Q}}})$ approach the real Q value function which will be used to guide the update of the ${\Theta ^\pi }$. The loss function of the critic network in an MSE sense is defined as
\begin{equation} \label{e30}
L({\Theta ^{\rm{Q}}}) = \frac{1}{{{X_2}}}\sum\nolimits_{i = 1}^{{X_2}} {{{\left({y_i} - Q({{\bf{s}}_i},{{{\bf{\tilde a}}}_i}|{\Theta ^{\rm{Q}}})\right)}^2}},
\end{equation}
where ${y_i} = {r_i} + \chi Q'({{{\bf{s'}}}_i},\pi '({{{\bf{s'}}}_i}|{\Theta ^{\pi '}})|{\Theta ^{{\rm{Q}}'}})$ can be seen as a label. However, the concept of label is quite different from that of the supervised learning. Here the label is not given by handcraft but determined by the output of the target critic as well as the environment reward feedback. Thus, ${\Theta ^{\rm{Q}}}$ can be updated by ${\nabla _{{\Theta ^{\rm{Q}}}}}L$ and $Q(s,a|{\Theta ^{\rm{Q}}})$ will gradually approach the real Q-value.
\begin{algorithm}[!t]
\caption{LSTM-DDPG approach training for our problem}
\label{alg2}
\small
\begin{algorithmic}[1]
 \STATE  \textbf{Initialize:} Initialize ${\Theta ^{\rm{L}}}$, ${\Theta ^\pi}$, ${\Theta ^{\rm{Q}}}$ and memory buffer $\Omega$.

 Copy the weights of online networks to the target networks: ${\Theta ^{\pi '}} \leftarrow {\Theta ^\pi },{\Theta ^{{\rm{Q}}'}} \leftarrow {\Theta ^{\rm{Q}}}$

\STATE Train the LSTM neural network using \eqref{e26}. Acquiring the convergent model ${\Theta ^{{{\rm{L}}^ * }}}$.
\STATE \textbf{For ${\rm {episode}} = 1, \cdots, {\cal T}$ do:}
\STATE \quad Initialize a random process ${\cal N}$ for the exploration.
\STATE \quad Initialize user request recorder ${\bf{R}}(0)$.
\STATE \quad Observe the initial state ${\bf{s}}(1)$
\STATE \quad \textbf{For $t = 1, \cdots, \Upsilon $ do:}
\STATE \quad \quad Obtain current viewpoint request ${V_k}(t)$. Select action\\
 \quad \quad ${\bf{\tilde a}}(t){\rm{ = }}\pi ({\bf{s}}(t)|{\Theta ^\pi }) + {\bf{n}}(t){{\rm{|}}_{{n_i}(t) \sim {\cal N}(0,\sigma _n^2)}}$
\STATE \quad \quad Execute action ${\bf{\tilde a}}(t)$, observe the instant reward $r(t)$
\STATE \quad \quad Obtain next viewpoint request ${V_k}(t+1)$
\STATE \quad \quad Update the request recorder ${\bf{R}}(t) \to {\bf{R}}(t + 1)$.
\STATE \quad \quad Obtain ${\bf{\hat p}}(t + 2)$ by \eqref{e25}.
\STATE \quad \quad Observe the new state ${\bf{s}}(t + 1)$ and vectorize it.
\STATE \quad \quad Store tracing point $({\bf{s}}(t),{\bf{\tilde a}}(t),r(t),{\bf{s}}(t + 1))$ in $\Omega $
\STATE \quad \quad Randomly sample a mini-batch of ${{X_2}}$ points\\
 \quad \quad $({\bf{s}}(t),{\bf{\tilde a}}(t),r(t),{\bf{s}}(t + 1))$ from $\Omega $.
\STATE \quad \quad Compute ${y_i} = {r_i} + \gamma Q'({{{\bf{s'}}}_i},\pi '({{{\bf{s'}}}_i}|{\Theta ^{\pi '}})|{\Theta ^{{\rm{Q}}'}})$
\STATE \quad \quad Update the online critic network ${{\Theta ^{\rm{Q}}}}$ by \eqref{e30}.
\STATE \quad \quad Update the online actor network ${{\Theta ^\pi }}$ by \eqref{e33}.
\STATE \quad \quad Update the target actor/critic network every $\psi $ steps:
\[\mathop {\rm{soft\;update}}\limits_{0 < \vartheta  < 1} \left\{ \begin{array}{l}
{\Theta ^{{\rm{Q}}'}} \leftarrow \vartheta  \cdot {\Theta ^{\rm{Q}}}{\rm{ + (1 - }}\vartheta {\rm{)}}{\Theta ^{{\rm{Q}}'}}\\
{\Theta ^{\pi '}} \leftarrow \vartheta  \cdot {\Theta ^\pi }{\rm{ + (1 - }}\vartheta {\rm{)}}{\Theta ^{\pi '}}
\end{array} \right.\]
\STATE \quad \textbf{End For}
\STATE \textbf{End For}
\end{algorithmic}
\end{algorithm}

The actor is aimed at producing a optimal policy which can acquire maximum cumulative reward, which is equivalent to maximum Q-value. ${y_i}$ is sent to the actor network as a current true Q-value. Then, for the current policy evaluation, a performance objective function is designed as
\begin{equation} \label{e31}
{J_\beta }(\pi ) = {E_{s \sim {\rho ^\beta }}}\left[Q(s,a|{\Theta ^{\rm{Q}}})\right],
\end{equation}
which estimates the expectation of $Q(s,a|{\Theta ^{\rm{Q}}})$ under the state distribution ${s \sim {\rho ^\beta }}$. Hence, the gradient in the view of ${\Theta ^\pi }$ is
\begin{equation} \label{e32}
\begin{array}{l}
{\nabla _{{\Theta ^\pi }}}{J_\beta }(\pi ) = \\ {E_{s \sim {\rho ^\beta }}}\left[{\nabla _a}Q(s,a|{\Theta ^{\rm{Q}}}){|_{a = \pi (s|{\Theta ^\pi })}} \cdot {\nabla _{{\Theta ^\pi }}}\pi (s|{\Theta ^\pi })\right].
\end{array}
\end{equation}
Monte Carlo method is adopted to estimate this expectation using the mini-batch samples with a size of $I$, which yields a unbiased estimation:
\begin{equation} \label{e33}
\begin{array}{l}
{\nabla _{{\Theta ^\pi }}}{J_\beta }(\pi ) \approx \\
\frac{1}{{{I}}}\sum\nolimits_i {\left({\nabla _a}Q({{\bf{s}}_i},{{{\bf{\tilde a}}}_i}|{\Theta ^{\rm{Q}}}){|_{{{{\bf{\tilde a}}}_i} = \pi ({{\bf{s}}_i}|{\Theta ^\pi }){\rm{ + }}{\bf{n}}(t)}} \cdot {\nabla _{{\Theta ^\pi }}}\pi ({{\bf{s}}_i}|{\Theta ^\pi })\right)}.
\end{array}
\end{equation}
We can see that ${{\Theta ^\pi }}$ is learned by following the direction of ${\nabla _{{\Theta ^\pi }}}{J_\beta }(\pi )$. 

As described in Algorithm \ref{alg2}, the target actor network ${\Theta ^{\pi '}}$ and target critic network ${{\Theta ^{{\rm{Q}}'}}}$ are soft-updated in every a few steps. 

The overall training process of the proposed scheme is also summarized in Algorithm \ref{alg2}. In Algorithm \ref{alg2}, ${\cal T}$ is the total training episodes and $\psi $ is the step interval between the target networks and the online networks in parameters update. $\vartheta $ is the coefficient of the soft-update which is normally set to 0.001. Note that, the proposed algorithm is executed centrally in the MEC server. Although different users has slightly different interests on the same content of VR video, the statistical trend of the viewpoints' popularity remains constant \cite{8728029}. Hence, the trained model can generally cater to different users without updates if the same VR video is delivered.

\section{NUMERICAL SIMULATIONS AND ANALYSES}
\label{sec5}

In this section, we evaluate the performance of the proposed scheme via numerical examples. In the simulation, we divide the 2D domain into a fixed number of 2D tiles with ${N_{{\rm{col}}}} = 7$ and ${N_{{\rm{row}}}} = 5$. For one viewpoint, $Z = 4$ with $N_{{\rm{col}}}^{{\rm{FOV}}} = 2$ and $N_{{\rm{row}}}^{{\rm{FOV}}} = 2$. In addition, for less shadow effect\cite{8552403}, we set both ${\Delta _{\rm{h}}}$ and ${\Delta _{\rm{v}}}$ equal to 1. Then, the number of viewpoints $K = 24$. The total size of all the 2D tiles is set to $Q = 5.25\;{\rm{Gbits}}$ \cite{Chen2019Data}, and the parameters of cache capacity are set to ${M_{\rm{L}}} = 3$ and ${M_{\rm{E}}} = 8$, respectively. Adam optimizer \cite{Kingma2014Adam} is used to learn the parameters ${\Theta ^\pi}$, ${\Theta ^{\rm{Q}}}$, and ${\Theta ^{\rm{L}}}$ with a learning rate of ${10^{ - 4}}$, ${10^{ - 4}}$ and ${10^{ - 5}}$ respectively. In addition, a $0.35$ dropout rate is used in the hidden layers to avoid overfitting. Other parameters are set by referring to the related works, for example \cite{8114362, 7412759}, which are listed in Table \ref{tab1}.

\vspace{-1em}
\begin{table}[!hpbt]
\centering
\caption{Simulation Parameters} \label{tab1}
\begin{tabular}{c|c||c|c}
\toprule[1pt]
    \textbf{Parameter} & \textbf{ Value} & \textbf{Parameter} & \textbf{ Value}\\
    \hline\hline
      ${P_B}$ &  $30\;\rm{dBm}$ & ${M_{\rm{E}}}$ & $8$ \\
    \hline
      $\sigma^2 $ & $ - 105\;\rm{dBm}$ & $\varphi$ & $3$ \\
    \hline
      ${R_{\rm{bh}}}$ & $10\;\rm{Gbit/s}$ & $\alpha $ & $2$ \\
     \hline
      ${f_L}$ & $3\;\rm{GHz}$ & $d$ & $100\;m$ \\
    \hline
      ${f_M}$ & $10\;\rm{GHz}$ & $\omega $ & $15\;\rm{cycle/bit}$\\
    \hline
      ${M_{\rm{L}}}$ &  $3$ & $\gamma $ & $0.85$\\
\bottomrule[1pt]
\end{tabular}
\end{table}
\vspace{-0.5em}

In order to imitate the real environment, the Zipf parameter space is set as ${\gamma _t} \in {\cal G} = \{ 0.7,1,1.5,2.5\}$ \cite{Sadeghi2017Optimal}, with a transition probability matrix ${\bf P}=\left[P_{ij}\right]_{i,j=0}^{4}$,
where $P_{ij}$ denotes the transition probability from ${\gamma ^j}$ to ${\gamma ^i}$. Note that, the transition probability matrix above is randomly generated and the entries can be any values, which has no impact on the system performance with our method. It should also be emphasized that the Zipf parameter space and the transition probability matrix are both unknown to our system during the simulated VR service process. Here they are just provided to demonstrate the operating mechanism of the simulation environment and do not involve in the problem solving.

Note that, prior to training the entire network, the LSTM network is pre-trained in advance. The experiment is run on an Intel Core i5-7500 CPU with $3.40\;{\rm{GHz}} \times 4$ cores. We provide the training statistics of the networks in Table \ref{tab2}.

\vspace{-1em}
\begin{table}[!hpbt]
\centering
\caption{Training Statistics} \label{tab2}
\begin{tabular}{c|c|c|c}
\toprule[1pt]
    \textbf{Training Phase} & \textbf{ Iterations} & \textbf{Total Time} &  \textbf{\tabincell{c}{Average Time \\(pre iteration)}}\\
    \hline\hline
      Pre-training &  1750 & $ 71\;\rm{s}$ & $0.0406\;\rm{s}$ \\
    \hline
      Training &  50000  & $ 775.867\;\rm{s}$ & $ 0.0155\;\rm{s}$ \\
    \hline
       Summation & 51750 & $ 846.867\;\rm{s}$ & $ 0.0164\;\rm{s}$ \\

\bottomrule[1pt]
\end{tabular}
\end{table}
\vspace{-1em}

\begin{figure*}[htbp]
    \centering
    \subfigure[]{
    \begin{minipage}[t]{0.33333\linewidth}
        \centering
        \includegraphics[width=1\columnwidth]{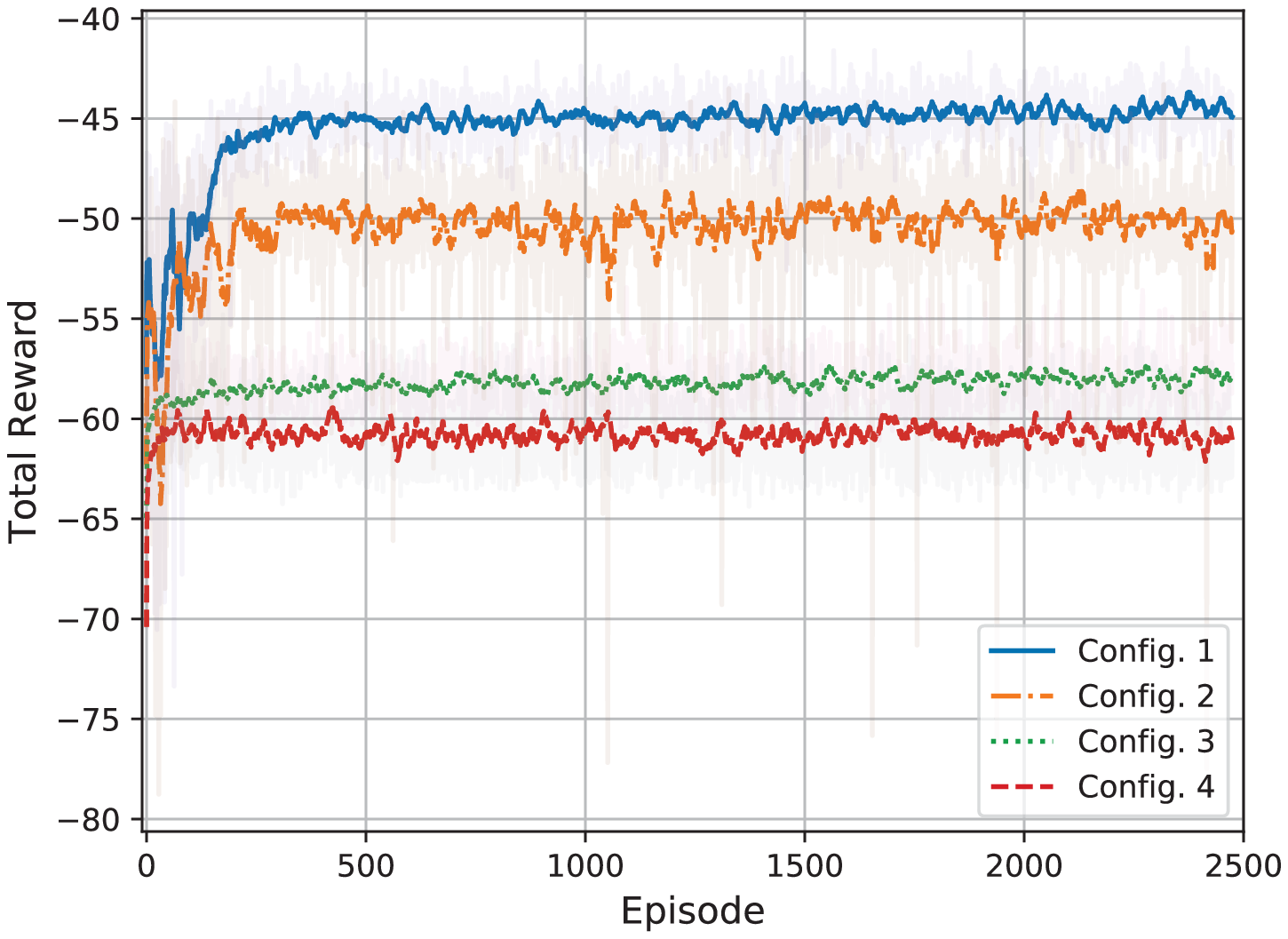}
    \end{minipage}%
    }%
    \subfigure[]{
    \begin{minipage}[t]{0.33333\linewidth}
        \centering
        \includegraphics[width=1\columnwidth]{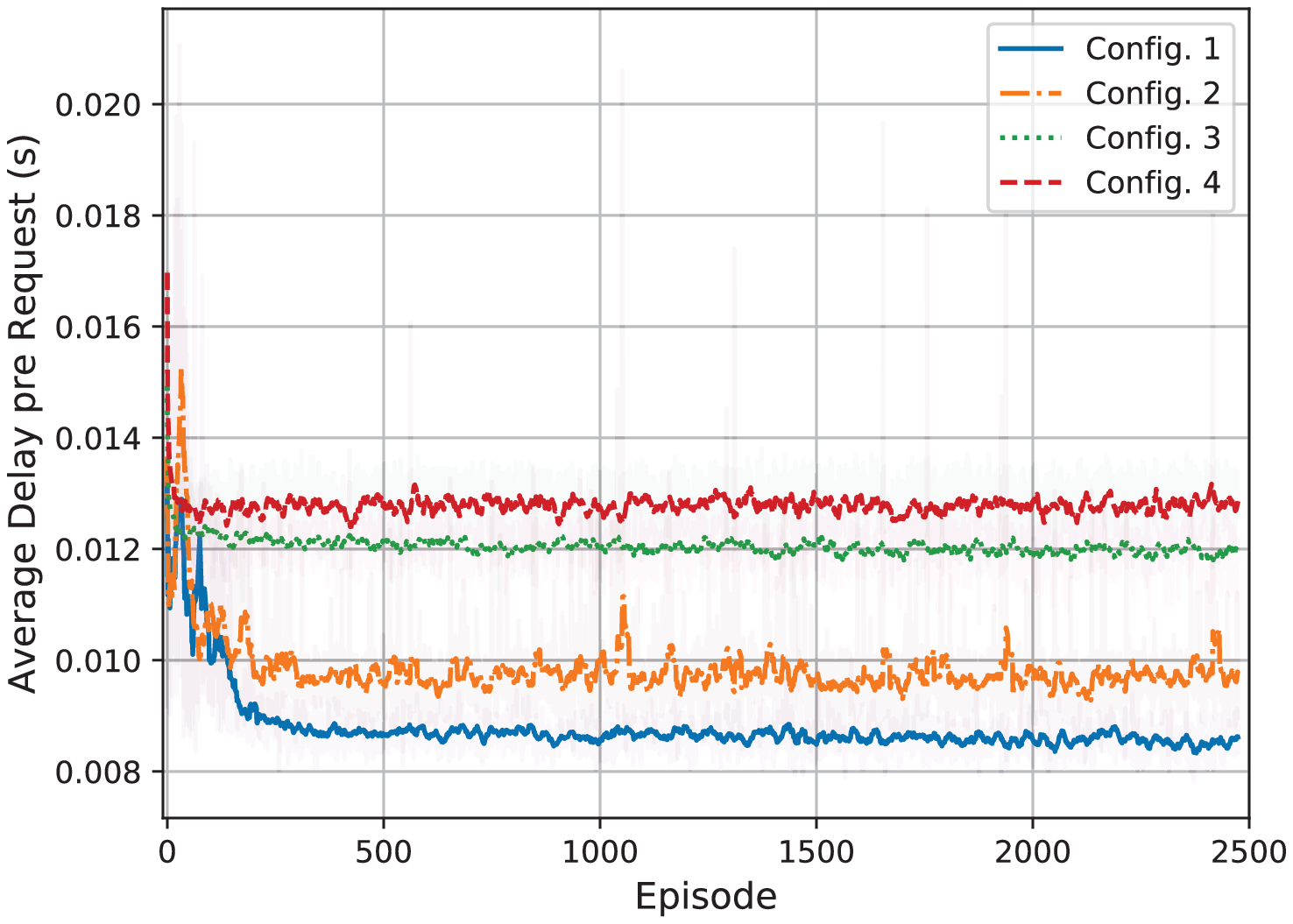}
    \end{minipage}%
    }%
    \subfigure[]{
    \begin{minipage}[t]{0.33333\linewidth}
        \centering
        \includegraphics[width=1\columnwidth]{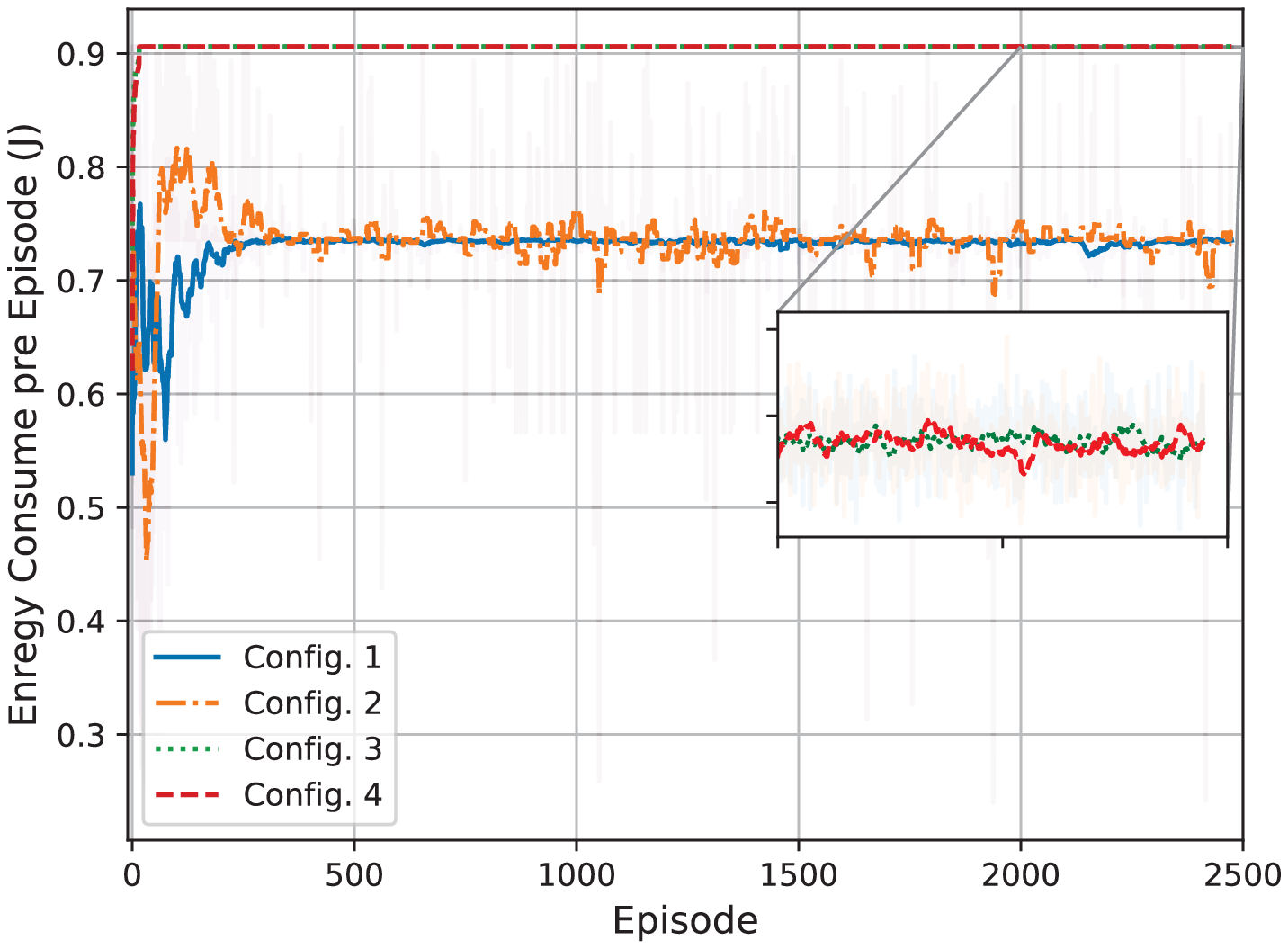}
    \end{minipage}
    }%
    \centering
    \caption{Performance comparison of different configurations of the proposed algorithm with respect to different metrics. (a) Total reward. (b) Average latency of each request. (c) Total energy consumption for one episode.}
    \label{experiment1}
\end{figure*}

\begin{figure*}[htbp]
    \centering
    \subfigure[]{
    \begin{minipage}[t]{0.33333\linewidth}
        \centering
        \includegraphics[width=1\columnwidth]{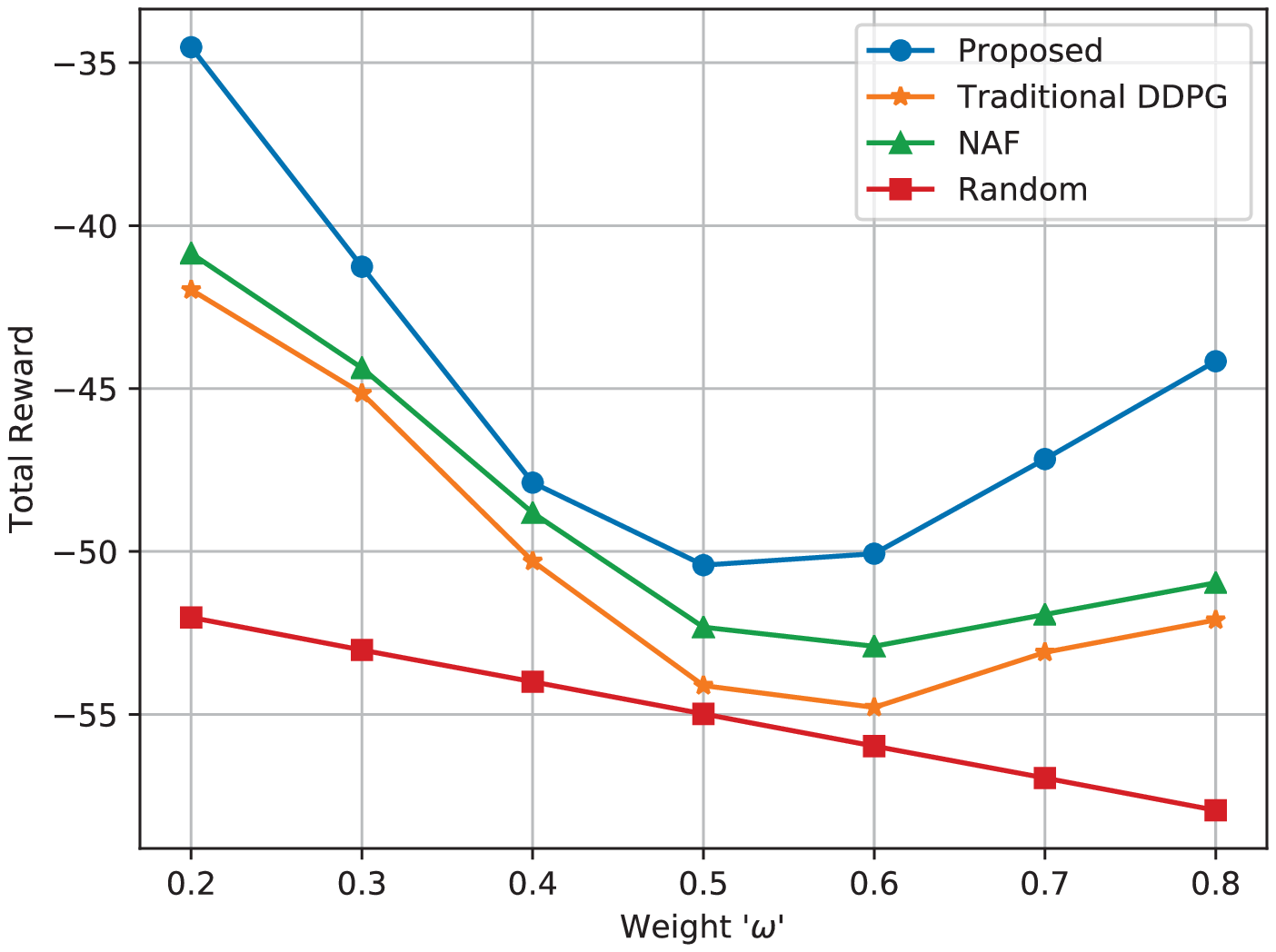}
    \end{minipage}%
    }%
    \subfigure[]{
    \begin{minipage}[t]{0.33333\linewidth}
        \centering
        \includegraphics[width=1\columnwidth]{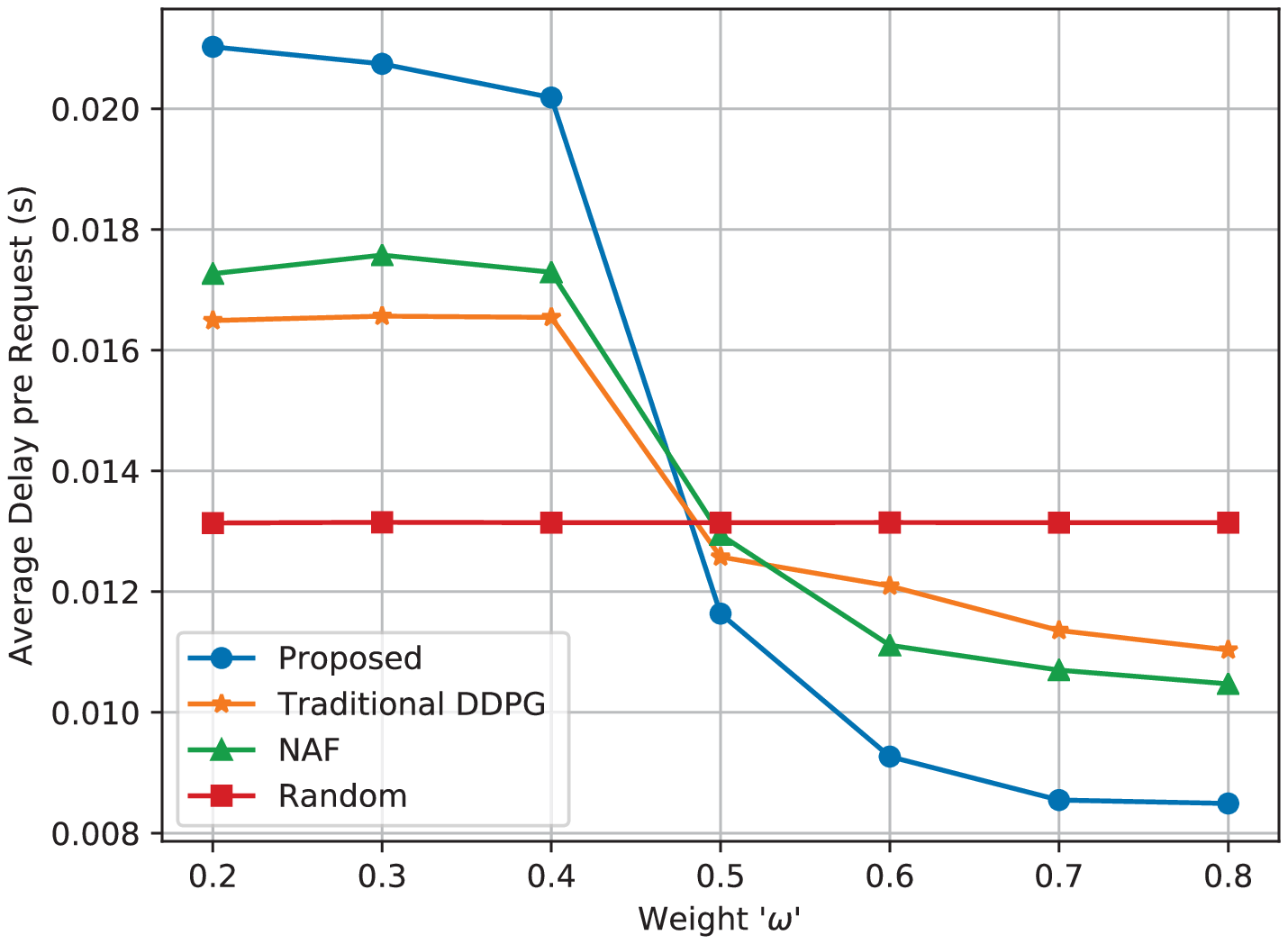}
    \end{minipage}%
    }%
    \subfigure[]{
    \begin{minipage}[t]{0.33333\linewidth}
        \centering
        \includegraphics[width=1\columnwidth]{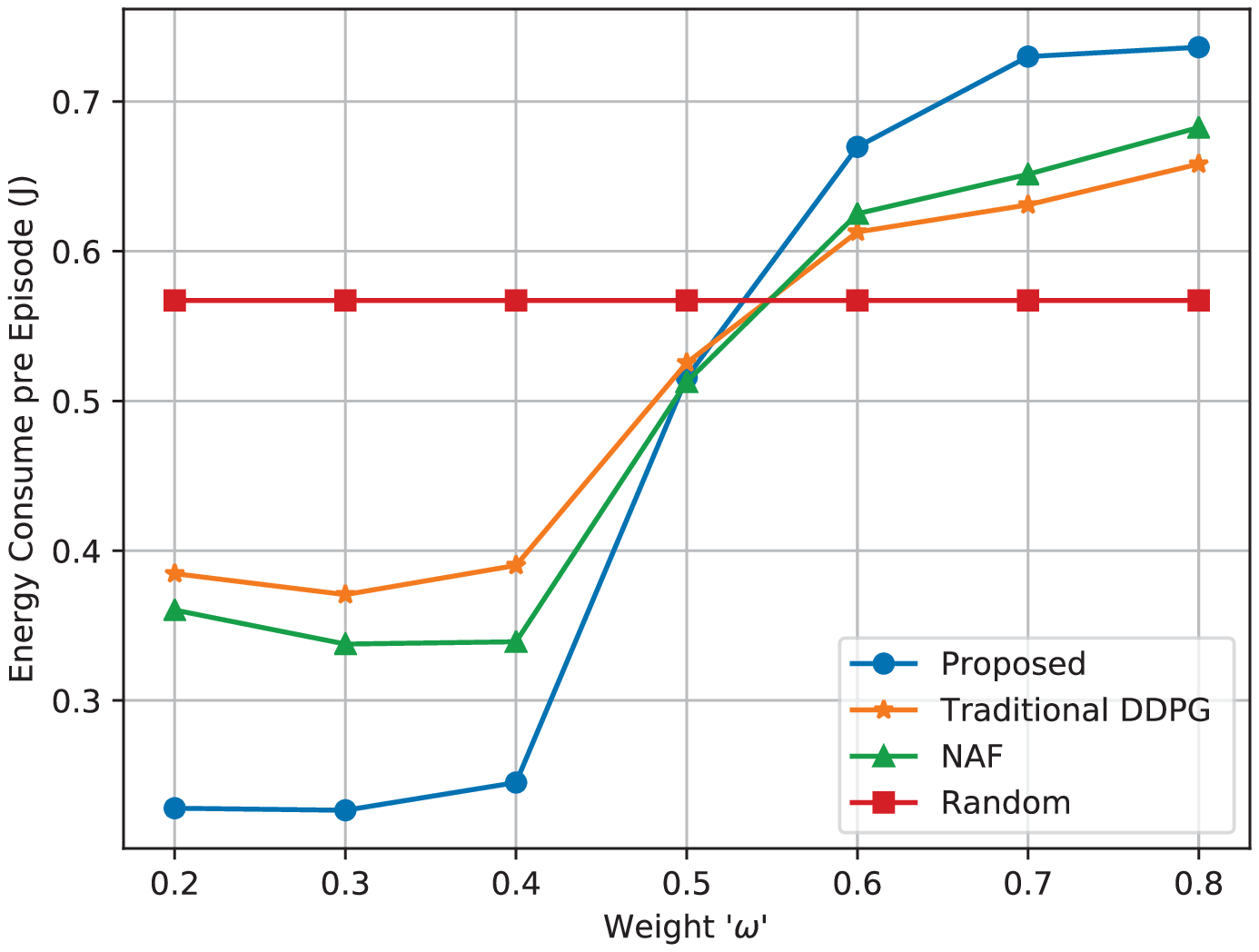}
    \end{minipage}
    }%
    \centering
    \caption{Impact of the weight $\omega$ on the system performance. (a) Total reward versus $\omega$. (b) Average latency of each request versus $\omega$. (c) Energy consumption versus $\omega$.}
    \label{experiment3}
\end{figure*}

\subsection{Baselines}
\subsubsection{Random Approach}
In this baseline approach, the policy is randomly formulated and a random action is executed regardless of the current state.

\subsubsection{Traditional DDPG Approach}
In this approach, the LSTM neural network is removed but the caching replacement and task segmentation are reserved. The traditional DDPG algorithm is used to solve the learning problem. Because of the absence of LSTM neural network, we replace the ${\bf{\hat p}}{\rm{(t + 1)}}$ with the ${\bf{R}}(t)$ in order to ensure the fairness of the input information compared with our method.

\subsubsection{Normalized Advantage Functions (NAF) Approach}
NAF algorithm \cite{DBLP:journals/corr/GuLSL16} is developed based on DQN and, similar to the DDPG algorithm, it is applicable to the high-dimensional or even continuous action control problem. But different from the DDPG, which uses two networks to output action and Q value, NAF integrates the action and Q value into one neural network by dividing the Q value into advantage term and state-value term. The caching replacement and deterministic task segmentation/offloading are also considered in implementing the NAF.

\subsection{Results and Discussions}

In Fig. \ref{experiment1}, we investigate the impacts of four different configurations of the proposed algorithm on the system performance, and we let $\omega = 0.8$ in the simulations:

\begin{itemize}
  \item Config. 1: Both the dynamic caching replacement and the deterministic task segmentation/offloading are enabled in both the local VR device and the MEC server. This configuration is the standard version of the proposed scheme and is expected to achieve a best performance.
  \item Config. 2: The deterministic task segmentation/offloading is enabled in both devices; while the dynamic caching replacement is disabled in both devices.
  \item Config. 3: The the dynamic caching replacement is enabled; while the deterministic task segmentation/offloading is disabled.
  \item Config. 4: Neither of deterministic task segmentation/offloading and the dynamic caching replacement is enabled.
\end{itemize}

\begin{figure*}[htbp]
    \centering
    \subfigure[]{
    \begin{minipage}[t]{0.386\linewidth}
        \centering
        \includegraphics[width=1\columnwidth]{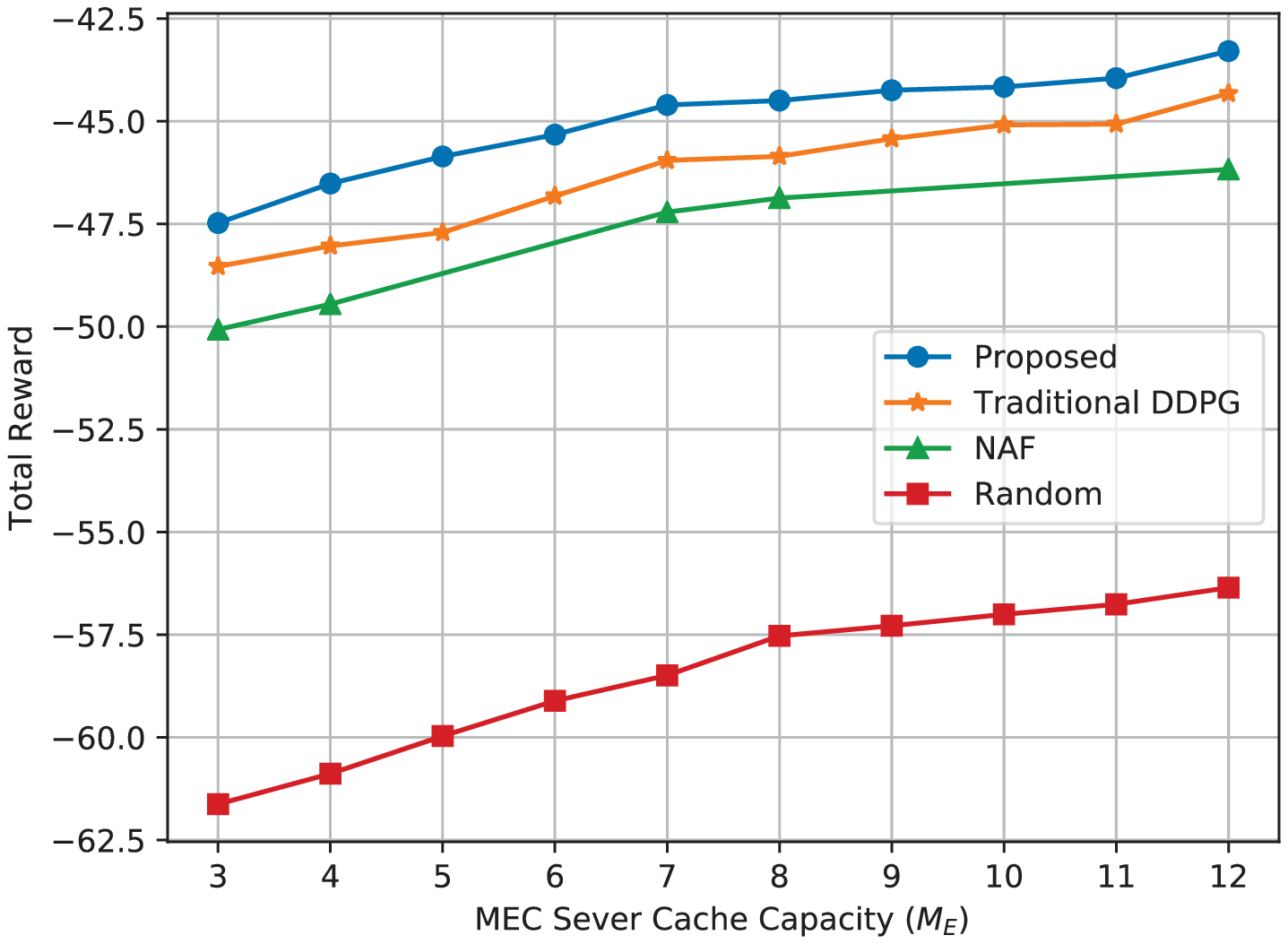}
    \end{minipage}%
    }%
    \subfigure[]{
    \begin{minipage}[t]{0.386\linewidth}
        \centering
        \includegraphics[width=1\columnwidth]{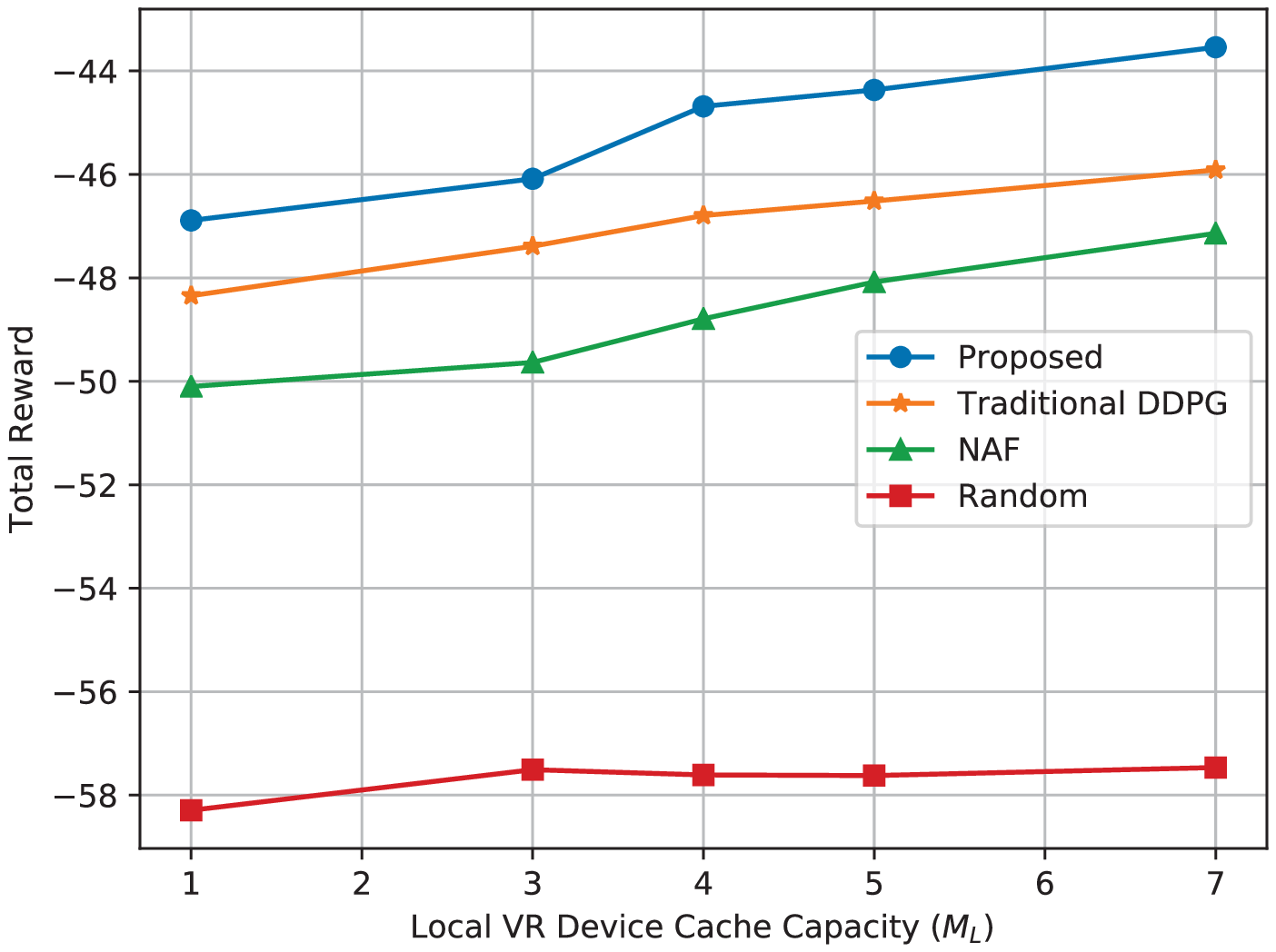}
    \end{minipage}%
    }%
    \centering
    \caption{Impact of the device cache capacities on the total reward. (a) Total reward versus the MEC server's cache capacity with ${M_{\rm{L}}} = 3$. (b) Total reward versus the local VR device's cache capacity with ${M_{\rm{E}}} = 8$.}
    \label{experiment2}
\end{figure*}

\begin{figure*}[htbp]
    \centering
    \subfigure[]{
    \begin{minipage}[t]{0.386\linewidth}
        \centering
        \includegraphics[width=1\columnwidth]{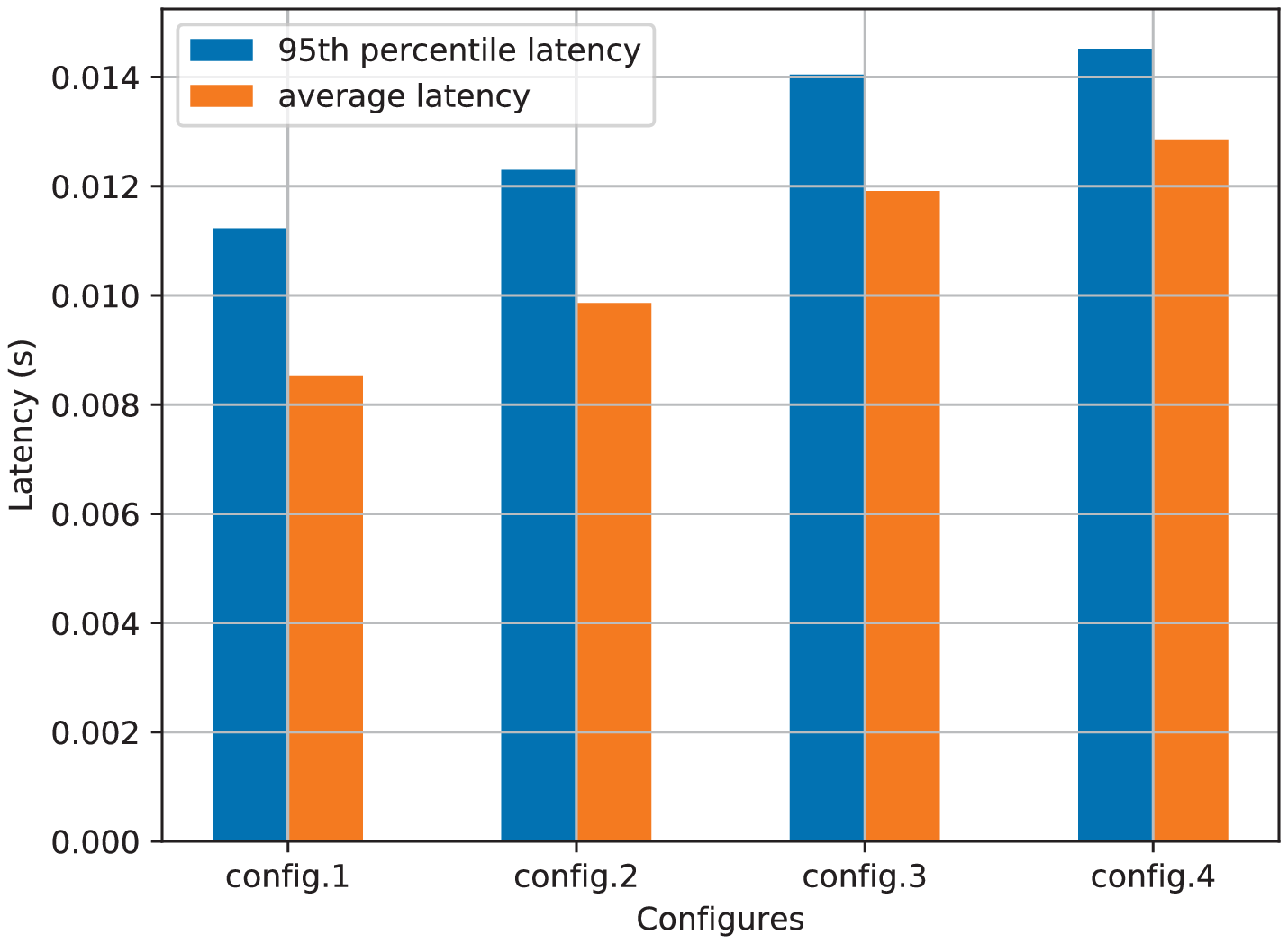}
    \end{minipage}%
    }%
    \subfigure[]{
    \begin{minipage}[t]{0.386\linewidth}
        \centering
        \includegraphics[width=1\columnwidth]{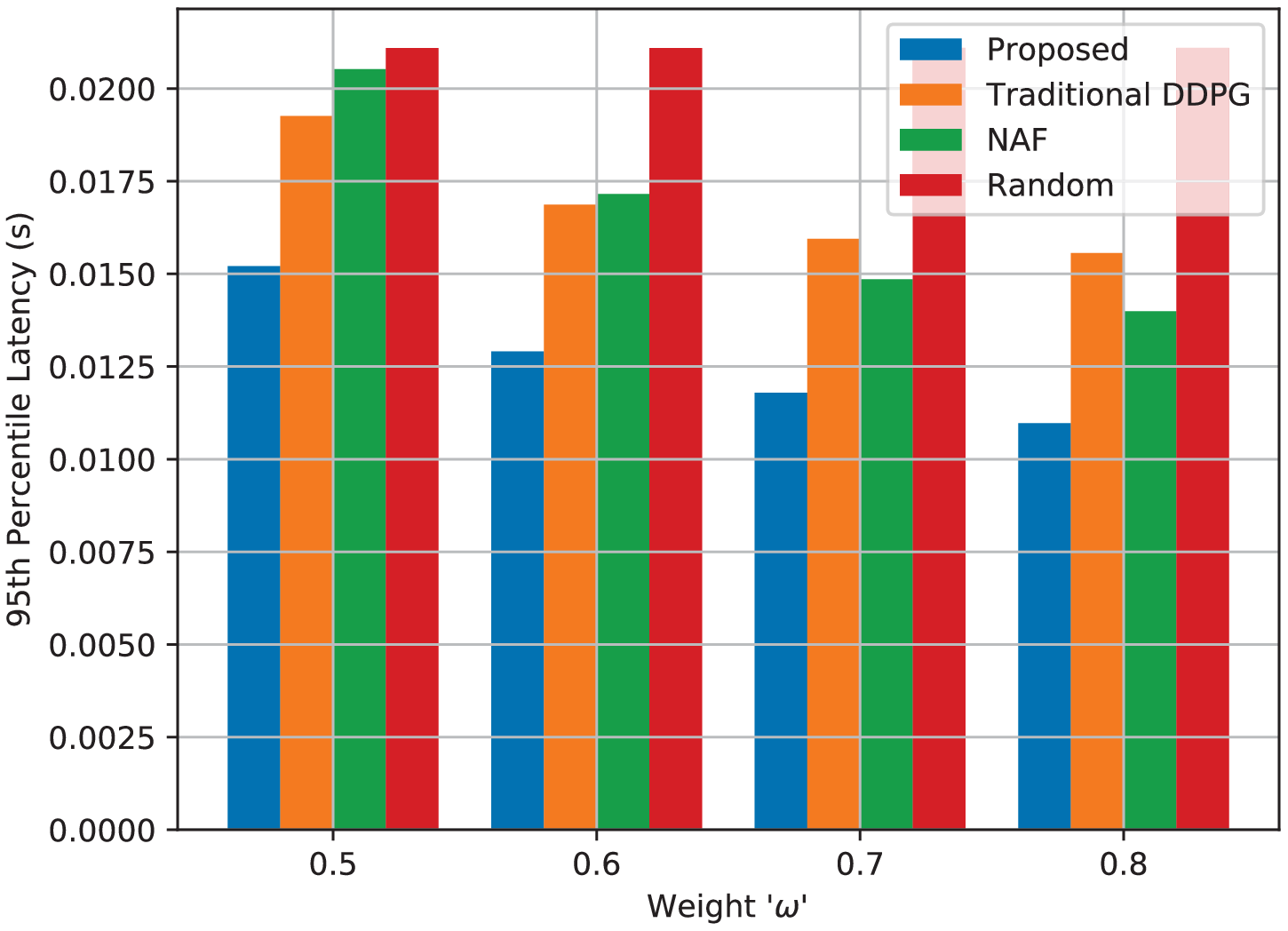}
    \end{minipage}
    }%
    \centering
    \caption{Performance comparison of different configurations and different algrithoms with respect to 95th percentile latency. (a) Different configurations. (b) Different weights.}
    \label{R3C1}
\end{figure*}

To guarantee fairness, the performance comparisons with the four different configurations share the same parameter setting as listed in Table~\ref{tab1}. From Fig. \ref{experiment1}(a), we observe that the standard Config. 1 yields maximum gain in terms of the total reward. In addition, Config. 2 is superior to Config. 3, which implies that the task segmentation/offloading module contributes more to gain than the dynamic caching replacement. This is due to the fact that the task segmentation concurrently mobilizes computing and caching resources of all devices in the system, while the caching replacement only leverages the caching resources to improve hit ratio. The performance gain is higher and more stable after adding the caching replacement, because the cache hit ratio in each slot is higher with the aid of the dynamic caching replacement owing to the field overlap. As a result, the transmitted date size is smaller and the fluctuation is lower. In terms of latency, we plot the average latency curves in Fig. \ref{experiment1}(b) which is equal to total latency of one episode divided by $T$. It can be observed that, the average latency of each task can decline and converges to 9 ms by adopting caching replacement and task segmentation with $\omega  = 0.8$. Fig. \ref{experiment1}(c) indicates that the proposed model maintains the lowest energy consumption after convergence. It also shows that the caching replacement has no effect on the amount of energy expenditure but still contributes to the stability. From the perspective of convergent behavior of the proposed approach, as shown in Fig. \ref{experiment1}, we find that the agent can earn a stabilized mean of reward after 2500 episodes, which means that the agent has acquired the knowledge of the entire system and the proposed algorithm gradually converges. Note that we set $\Upsilon = 100$ in the simulation and, thus, one episode corresponds to $100$ iterations according to the Algorithm \ref{alg2}.

The performance comparisons between the proposed method and all the baseline algorithms versus the weight $\omega$ are presented in Fig. \ref{experiment3}. It can be seen from Fig. \ref{experiment3}(a) that the proposed algorithm outperforms all the baselines regardless the value of $\omega$. Note that a notch can be identified on each of the total reward curves. This is due to the tradeoff between the latency and the energy consumption. The larger the weight is, the more important the latency is in the system. When the weight is around 0.5, the agent choose to make actions that ensures the equivalent significance of these two considerations. From Figs. \ref{experiment3}(b) and (c), we can see that, the average latency decreases with the increase of $\omega$, at the expense of the system's energy consumption. This also reveals the tradeoff between the two considerations. On the other hand, in Figs. \ref{experiment3}(b) and (c), we can find that, when $\omega$ is at a higher value which means the latency is paid more attention to (e.g. the system is fully charged) the proposed algorithm consumes more energy to ensure a lower latency than the other three algorithm. This demonstrates that the proposed approach is more sensitive to the weight variation and more flexible to the different emphasis of the system. We also readily observe from Figs. \ref{experiment3}(b) and (c) that, the random approach has no sense of the weight variation, which is reasonable since random policies and actions are selected and executed.

Fig. \ref{experiment2} illustrates the impact of the cache capacity at MEC server and local VR device. We investigate the relationship between the total reward and the variations of the cache capacities ${M_{\rm{L}}}$ and ${M_{\rm{E}}}$, and the other system parameters remain unchanged. When the MEC server's cache capacity is restricted to $8\tau$, we change the local VR device's capacity from $\tau$ bit to $7\tau $ bit in order to satisfy the inequality ${M_{\rm{L}}} \le {M_{\rm{E}}}$. Similarly, we change the MEC server's cache capacity from $3\tau $ bit to $12\tau $ bit and limit the cache capacity of the local VR device to $3\tau $. Both Figs. \ref{experiment2}(a) and (b) suggest that the proposed algorithm outperforms all the other baseline methods in terms of total reward performance and the total reward of all considered algorithms increases with the cache capacity. This is because that the MEC server or the local VR playback device can cache more content with the improvement of the cache capacity such that the agent is able to reduce the latency with the same energy consumption. Fig. \ref{experiment2}(b) shows that the advantage of the proposed method becomes more significant as the cache capacity of the local device increases. This implies that the cache resource at local VR playback device is very helpful for the service, especially when the specialized VR playback device with larger cache capacity is used. However, it is noted that the curves in Fig. \ref{experiment2}(b) grows slightly slower than those of Fig. \ref{experiment2}(a). The reason is that the computing latency and energy dissipation of the local VR device is larger than that of the MEC server, which reduces the gain from the cache capacity improvement. We further evaluate the performance of the proposed models and algorithm with respect to 95th percentile latency during the consecutive $T$ requests, which is presented in Fig. \ref{R3C1}. From Fig. \ref{R3C1}(a), we again confirm that the Config. 1 significantly outperforms the other configurations. We also readily observe from Fig. \ref{R3C1}(b) that, the proposed algorithm outperforms all the other baseline methods with respect to 95th percentile latency with different weight $\omega$.

\section{Conclusion}\label{sec6}

VR is among the killer applications of the future wireless networks, which will offer unprecedented experiences and possibilities. In this paper, we consider realizing wireless VR video service by exploiting a MEC-assisted network. The problem of deterministic offload and dynamic caching replacement are first jointly examined both at the MEC server and the local VR device. A hybrid policy is formulated to minimize the system latency and the energy consumption as well as seek a trade-off between them. To solve the yielded challenging multi-objective optimization problem and overcome the uncertainty of the viewpoint popularity, a mode-free MDP is established and an LSTM-DDPG algorithm is proposed to learn the optimal policy. The superior performance of the proposed scheme compared to the baseline methods and the gain contributed by each module are confirmed by the numerical simulations. Our future work will concentrate on more complicated scenarios such as multiple wireless VR users and multiple MEC nodes, where schemes for transmission resource management and interference suppression must be investigated.

\ifCLASSOPTIONcaptionsoff
  \newpage
\fi

\balance

\end{document}